\documentclass{llncs}
\usepackage{bbding}

\usepackage{dsfont,mathtools,multicol,float,subcaption,eufrak,amssymb,listings,hyperref,underscore}
\usepackage{thm-restate}
\newcommand{\qedhere}{\qed}
\usepackage{tikz}
\usepackage{cleveref}
\usepackage[inline]{enumitem}

\usetikzlibrary {arrows.meta,graphs,quotes}
\usetikzlibrary{arrows,trees,backgrounds,automata,shapes,plotmarks,decorations,fit}
\tikzstyle{block}=[rectangle,draw, thin, inner sep=3pt, text centered,fill=orange!20!yellow!20] %drop shadow
\tikzstyle{net}=[draw,cloud,fill=yellow!20,aspect=3,inner sep=1pt]%drop shadow
\tikzstyle{dev}=[draw,circle,fill=yellow!20,aspect=2,inner sep=1pt,minimum size=.6cm]%drop shadow
\tikzstyle{pre}=[<-,shorten <=1pt,>=stealth']
\tikzstyle{post}=[->,shorten >=1pt,>=stealth']
\tikzstyle{bi}=[<->,shorten >=1pt,shorten <=1pt, >=stealth']
\tikzstyle{every initial by arrow}=[initial text={},initial distance=1em,post]
\tikzstyle{every state}=[minimum size=0.6cm,fill=cyan!20!yellow!20]%drop shadow
\tikzstyle{transition}= [post,shorten >=1pt,node distance=2cm, inner sep=2pt,bend angle=20]

\spnewtheorem*{MP}{Main Problem}{\bfseries}{\itshape}
\spnewtheorem{thm}{Theorem}{\bfseries}{\itshape}
\spnewtheorem{con}[theorem]{Construction}{\bfseries}{\itshape}
\crefname{thm}{Thm.}{Thms.}
\crefname{con}{Constr.}{Constrs.}
\crefname{proposition}{Prop.}{Props.}
\crefname{lemma}{Lem.}{Lemmas}
\crefname{definition}{Def.}{Defs.}
\crefname{corollary}{Cor.}{Cors.}
\crefname{figure}{Fig.}{Figs.}
\crefname{section}{Sect.}{Sections}
\crefname{equation}{Eq.}{Eqs.}

\newcommand{\NoZ}{\mathop{\nu}}
\newcommand{\ent}{\mathcal{H}}
\newcommand{\capa}{\mathcal{C}}
\newcommand{\band}{\mathcal{B}}
\newcommand{\bandc}{\mathcal{BC}}
\newcommand{\bandh}{\mathcal{BH}}
\newcommand{\gro}{\mathcal{G}}
\newcommand{\trans}[1]{\xrightarrow{#1}}
\newcommand{\Trans}[1]{\cdot\!\cdot\!\!\xrightarrow{\!\!#1}}
\newcommand{\pset}[1]{\mathcal{P}(#1)}
\newcommand{\Pset}[1]{\mathcal{P}_+(#1)}
\newcommand{\lett}[1]{\ell(#1)}

\newcommand{\guard}{\mathfrak{g}}
\newcommand{\reset}{\mathfrak{r}}

\newcommand{\abs}{\mathrm{abs}}

\newcommand{\full}{\mathrm{full}}
\newcommand{\squeeze}{\mathcal{S}}
\newcommand{\Net}{\mathsf{Net}}
\newcommand{\Sep}{\mathsf{Sep}}

\newcommand{\x}{\mathbf{x}}
\newcommand{\y}{\mathbf{y}}

\DeclareMathOperator{\supp}{\mathrm{supp}}
\DeclareMathOperator{\rad}{\mathrm{rad}}
\DeclareMathOperator{\resets}{\mathrm{resets}}
\newcommand{\Path}{\mathop{path}}

\newcommand{\Word}{\mathop{word}}
\newcommand{\Let}{\mathop{letters}}

\newcommand{\length}{\mathop{\mathrm{dur}}}

\newcommand{\D}{\mathcal{D}}
\newcommand{\Dset}{\mathds{D}}

\newcommand{\pqd}{\prescript{p}{q}\! \mathcal{D}}
\newcommand{\pql}{\prescript{p}{q}\! L}
\newcommand{\rfl}{\prescript{r}{f}\! L}
\newcommand{\rrl}{\prescript{r}{r}\! L}

\newcommand{\real}{\mathds{R}}
\newcommand{\rat}{\mathds{Q}}
\newcommand{\nat}{\mathds{N}}
\newcommand{\Log}{\mathds{L}}

\newcommand{\aut}{\mathcal{A}}

\newcommand{\eqapprox}{\mathrel{\raisebox{-0.3ex}{$\overline{\phantom{\hspace{-1.3mm}}\approx}$}}}

\newcommand{\dr}{\overrightarrow{d}}

\newcommand{\aaa}{node[red]{$\circ$}}
\newcommand{\bbb}{node[red]{$\circ$}}

\newcommand{\0}{\mathbf{0}}
\newcommand{\1}{\mathbf{1}}

\newcommand{\RTA}{\textsc{RsTA}}
\newcommand{\CTA}{\textsc{SFA}}

\begin{document}
\title{Weighing Obese Timed Languages\thanks{This is an expanded version (with proofs) of an article  published  by Springer Nature in the Proceedings of the 19th International Conference on
Reachability Problems --- RP 2025. This work was funded by ANR project MAVeriQ ANR-CE25-0012.}}
\author{Eugene Asarin\orcidID{0000-0001-7983-2202}\inst{1}\Envelope\and Aldric Degorre\orcidID{0000-0003-2712-4954}\inst{1}\and C\u at\u alin Dima\orcidID{0000-0001-5981-4533}\inst{2}\and Bernardo Jacobo Inclán\orcidID{0009-0009-5323-7945}\inst{1} }

\institute{Université Paris Cité, CNRS, IRIF, Paris, France \email{\{asarin,adegorre,jacoboinclan\}@irif.fr} \and LACL, Université Paris-Est Créteil, France \email{dima@u-pec.fr}}
\maketitle
\begin{abstract}
%In previous works, the notion of bandwidth of a timed language characterizing the quantity of information per time unit
%(with a finite observation precision $\varepsilon$) has been introduced. Afterward,  timed regular languages have been classified 
%into three classes with respect to the asymptotic behavior of the bandwidth as $\varepsilon \to 0$: meager, normal, and obese. 
% In previous works, we introduced the notion of bandwidth of a timed language, characterizing the quantity of information per time unit (with a finite observation precision $\varepsilon$) and classified timed regular languages into three classes with respect to the asymptotic behavior of the bandwidth as $\varepsilon \to 0$: meager, normal, and obese. 
The bandwidth of a timed language characterizes the quantity of information per time unit (with a finite observation precision $\varepsilon$). Obese timed automata have an unbounded frequency of events and produce information at the maximal possible rate. In this article, we compute the bandwidth of any such automaton in the form $\approx\alpha/\varepsilon$. 
%
% In this article, we address the problem of computing the bandwidth of a class of timed automata which produce information with large rates per second. 
% These automata have been called \emph{obese} in \cite{3classes} 
% where their bandwidth under precision $\varepsilon$ was shown to be $\approx\alpha/\varepsilon$. 
%
Our approach
%includes several transformations of the automaton, with an intermediary step where 
%we compute the 
reduces the problem to computing the best reward-to-time ratio in a weighted timed graph
constructed from the given timed automaton, with weights corresponding to the entropy of auxiliary finite automata.
% An important step in the construction of this weighted timed automaton separates time-bounded components in the original automaton where 
% the high-rate information production is concentrated, and computes the growth rate of the timed language of each such component,
% %as the logarithm of an algebraic number,
% using the technique for computing the entropy of finite automata. 
% The weights employed in the construction of the weighted timed automaton are the growth rates of components. 
%

\end{abstract}

\section{Introduction}
  One of the first articles on automata \cite{ChomskyMiller} showed that the number of words of length $n$ in a regular language asymptotically grows as $2^{nH}$, and computed the rate $H$; also the rate of polynomial growth (for the case when $H=0$) was fully characterized in \cite{poly}. Later, the growth (exponential, polynomial, and even intermediary)  has been explored for context-free languages \cite{growth1,growth3}, finitely generated groups \cite{grigorchuk}, and in different terms (entropy rate of subshifts) in symbolic dynamics  \cite{Marcus}. Growth analysis of automata can be seen as  a quantitative version of reachability: counting the number of paths instead of verifying their existence.

An  important  motivation for studies of growth rate consists in its interpretation as information content (entropy, capacity). In the 40s, \cite{shannon48} related word counts to the  capacity of a discrete noiseless channel; in the 60s, \cite{3approaches} conceptualized the counting (``combinatorial'') approach to the quantity of information, as compared to the probabilistic and the algorithmic ones. According to this approach,  an element of a finite set $S$ conveys $\log_2 \#S$ bits of information, and in most cases, asymptotics of this amount w.r.t.~some size parameter is studied. Nowadays, counting-based analysis of  formal languages  provides a background to the theory of codes \cite{codes-automata},   and  to %practical
protocols \cite[Chapter 6]{imminkBook} implemented in every hard disk drive and DVD \cite{efmplus}. In a nutshell, one can encode a source language  with a growth rate $H$  into a channel with growth rate $H'$ only if  $H\leq H'$. Also, words of such a language can be ``zipped'' with a compression rate at most $H$. %(or equivalently, with slowdown $A/B$). 

Timed languages \cite{AD}, widely used to represent behaviors of real-time and cyber-physical systems, extend usual formal languages with timing information of each event.  In the long run, we aim to port the growth rate/information content analysis to timed languages and automata with theoretical and practical ambitions (e. g.~coding and compression). For 
%growth rate (or 
information quantity %) 
\textbf{per event}, we have defined and computed the growth rate in  \cite{entroJourn,timedCoding}. However, we believe that 
%both conceptually and 
practically it would be more relevant to consider growth with respect \textbf{to time}, not to the number of events.  Our main challenge can thus  be stated as follows: 
\begin{equation}\label{challenge}
\begin{minipage}{0.934\textwidth}\small\it
for a timed automaton,  define and compute  the \textbf{amount of information in bits per time unit} conveyed by its accepted words (observed with a precision $\varepsilon>0$). 
\end{minipage}
\end{equation}
Previously, we made several steps toward the solution of \eqref{challenge}. %, in this paper we make another, penultimate step. 
In \cite{bw-our}, we formalize the notion of amount of information for timed languages, coin the term \emph{bandwidth} for it, and prove that it  is relevant to constrained channel coding with bounded delay. The timed language being continuously infinite, even when the time is bounded, its elements cannot be counted. Instead, we follow the approach from \cite{kolmoEpsilon} and count the elements of an $\varepsilon$-net, approximating the timed words with a finite precision $\varepsilon>0$, and use this count to define the bandwidth. 

In \cite{3classes},  we classify languages of timed automata into three classes characterized by structural properties of the automata, that  differ by the way that information is conveyed (see the examples on \cref{fig:3kinds}). 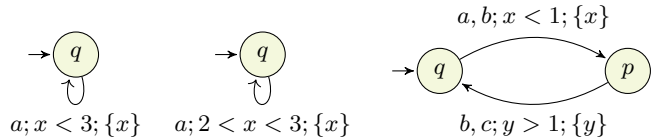
\begin{figure}[t]\small
\begin{center}
\begin{tikzpicture}
\node[state](q)[initial] {$q$};
\draw  (q) edge [loop below] node {$a;x<3;\{x\}$} (q);
\end{tikzpicture}
\ 
\begin{tikzpicture}
\node[state](q)[initial] {$q$};
\draw  (q) edge [loop below] node {$a;2<x<3;\{x\}$} (q);
\end{tikzpicture}
\ 
\begin{tikzpicture}
\node[state](q)[initial]   {$q$};
\node[state,right of=q,node distance=2.5cm](p) {$p$};
\draw [post] (q) edge [bend left, above] node {$a,b;x<1;\{x\}$} (p);
\draw [post]  (p) edge [bend left,below] node {$b,c;y>1;\{y\}$} (q);
\end{tikzpicture}
% \ 
% \begin{tikzpicture}[every node/.style={text=black}]
% \node[state,initial](p)  at (0,2.2)  {$p$};
% \node[state](q) at (2.5,2.2) {$q$};
% \node[state](r) at (1.25,0) {$r$};
% \draw [post] (p) edge [bend left,above] node[near start] {$x,y\in(0,1)$} node[below]{$a,b,\{x\}$}  (q);
% \draw [post] (q) edge [bend left=8mm,below] node {$x,y\in(0,1)$}node[above]{$c$} (p);
% \draw [post] (q) edge [below,sloped] node {$b,1<x<2$} (r);
% \draw [post] (r) edge [below,sloped] node {$6<x<7$} node[above,sloped,near start]{$b,\{xy\}$}(p);
% \draw  (q) edge [loop above] node {$0<x,y<1$} node[right=1mm]{$b,c$} (q);
% \draw  (r) edge [loop below] node {$a,b,c,1<x<2$} (r);
% \end{tikzpicture}
% \qquad
% \begin{tikzpicture}
% \node[state](q)[initial] {$q$};
% \draw  (q) edge [loop above] node {$a,b,x<3$} (q);
% \end{tikzpicture}
%   \begin{tikzpicture}
%       \node[state] (q1) at (2.0,0) {$q_1$};
% %      \node[state] (q2) at (2.0,-2.5) {$q_2$};
%       \node[state] (q0) [initial] at (-2.0,0) {$q_0$};
%       \draw [post] (q0) [in=110,out=70] edge node [below] {$\begin{matrix}a, x\leq 2\\x:=0\end{matrix}$}  (q1);
%       \draw [post] (q1) [in=-70,out=250] edge node [above] {$\begin{matrix}b, x\leq 1\\x:=0\end{matrix}$}  (q0);
%     \end{tikzpicture}
\end{center}
\caption{three timed automata and their asymptotic bandwidths (when $\varepsilon$ is small), respectively
%$\eqapprox
$\frac 1 \varepsilon$ (obese);  $\frac 1 2 \log(\frac 1 \varepsilon)$ (normal); and $2$ (meager). %Right: our running example (obese), bandwidth to be computed.%; and $0$ (meager and bounded
\label{fig:3kinds}}
\end{figure}
In \textbf{meager} automata, the bandwidth is $O(1)$. Information mostly comes from discrete choices, since all timed transitions are very constrained. In  \cite{CIAA}, we  characterized the exact value of the bandwidth of meager automata as the spectral radius of a finite-state abstraction.
In \textbf{normal} ones, the bandwidth is $\alpha \log 1/\varepsilon$, and the information is mostly conveyed  through  free choices of some duration every couple of time units.   In typical cases, the coefficient $\alpha$ is the number of ``degrees of freedom'' (continuous duration choices) per time unit in some optimal cycle. However, the question of computing $\alpha$ for the most general case remains open.
\textbf{Obese} automata have a bandwidth $\Omega(1/\varepsilon)$. In  some parts of such automata,  events can happen with  high frequency (without a lower bound on delays),  yielding a huge bandwidth.
Since very different bandwidth production mechanisms come into play in the three cases, bandwidth computation techniques also differ a lot.

In this paper,  we concentrate on obese automata and  compute the main asymptotic term of their bandwidth in the form  $\alpha/\varepsilon$. We proceed as follows:  after a preprocessing, we identify within the automaton all ``spots'' %, that is, strongly connected components 
that admit very fast information production, and compute the information rate $\alpha_\D$ of each such spot $\D$ (as the logarithm of the spectral radius of a matrix). Next, we transform the obese automaton into an ``abstract'' weighted timed graph, with states corresponding to spots, 
weights corresponding to their rates  $\alpha_\D$, and transitions to switching modes and resetting some clocks. Finally, the overall rate $\alpha$ corresponds to the maximal reward-per-time ratio in the abstract timed graph, computable due to \cite{alive}.
We prove that $\alpha/\varepsilon$ is indeed an upper and a lower bound for the bandwidth, 
and that $\alpha$ is the logarithm of an algebraic number. %, computed as the rational combination of several $\alpha_D$ of spots along the optimal cycle in the abstract graph. 

%We have thus solved the  problem \eqref{challenge} for the second  out of three cases. 

%We also believe that the obese timed automata are of a practical interest, as 
%illustrated by the  toy example of 

An example of a planetary robot modeled with an obese  timed automaton is provided in  \cref{fig:robot}. 
The robot records all events, observed with time granularity $\varepsilon$,  in a log file. 
The bandwidth $\alpha/\varepsilon$ corresponds to the disk space  (in bits per hour of the robot's life) required to store the log file.   \begin{figure}[t]
\begin{subfigure}[c]{.5\textwidth}
\begin{tikzpicture}\scriptsize
% \node at (-2,5){All transitions have guards: $x\leq 20,t\leq 25$};
% \node at (-2,4.5){All states have self-loops: $m,t=25,\{t\}$};
\node[state,rectangle,rounded corners](c) at (-3,1)   {cave};
\node[state,rectangle,rounded corners](h) at (-3,3)  {hill};
\node[state,rectangle,rounded corners](u) at (-5,2)  {up};
\node[state,rectangle,rounded corners](d) at (-1,2)  {down};
\node[state,rectangle,rounded corners](ch) at (-3,5.5)  {charge};

 \draw [post] (c) edge [above,sloped] node {$\uparrow\!\! u;\{y\}$}   (u);
 \draw [post] (u) edge [below,sloped] node {$\downarrow \!\!u;y\in[2,3]$}   (h);
 \draw [post] (h) edge [sloped, below] node {$\uparrow\!\! d;\{y\}$}   (d);
 \draw [post] (d) edge [above,sloped] node {$\downarrow \!\!d;y\in[3,4]$}   (c);

\draw [post] (u) edge [bend left=1.6cm,sloped,above] node {$\uparrow\!\! c;t\in [8,12];\{y\}$}    (ch);
\draw [post] (h) edge [sloped,above, bend left] node {$\uparrow\!\! c;t\in [8,12];\{y\}$}   (ch);

\draw [post] (ch) edge [sloped, above, bend left=1.6cm] node {$\downarrow \!\!c;y\in[7,8];\{x,y\}$}   (d);
\draw [post] (ch) edge [sloped, above, bend left] node {$\downarrow \!\!c;y\in[7,8];\{x\}$}   (h);

% \draw  (c) edge [loop left] node {$m,t=25,\{t\}$} (c);
% \draw  (u) edge [loop left] node {$m,t=25,\{t\}$} (u);
% \draw  (d) edge [loop right] node {$m,t=25,\{t\}$} (d);
% \draw  (h) edge [loop above] node {$m,t=25,\{t\}$} (h);
 %\draw  (ch) edge [loop above] node {$t=25,\{t\}$} (ch);
 \draw  (c) edge [loop right,red] node[black] {$a,b,e,f$} (c);
 \draw  (h) edge [loop left,red] node[black] {$a,b$} (h);
\end{tikzpicture}
\end{subfigure}%
\begin{subfigure}[c]{.5\textwidth}\small
All transitions have guards: $x\leq 20,t\leq 25$;\\
All states have self-loops: $m,t=25,\{t\}$.\\\\
\begin{tabular}{|l|l|}
\hline
\multicolumn{2}{|c|}{clocks}\\
\hline
$x$ & battery charge $=20-x$\\
$t$ & daytime clock\\
$y$ & time elapsed in current action\\\hline
\multicolumn{2}{|c|}{events}\\
\hline
$a,b,e,f$ & exploration events\\
$\uparrow\!\! u,\downarrow \!\!u$ & start and end going uphill\\
$\uparrow\!\! d,\downarrow \!\!d$ & start and end going downhill\\
$\uparrow\!\! c,\downarrow \!\!c$ & start and end charging\\
$m$ & midnight\\\hline
\end{tabular}
\end{subfigure}

\caption{\small A Martian robot modeled by an obese TA. The robot explores two spots: a cave and a hill, can travel between them, and must charge the batteries during daytime (starting from 8 to 12, for 7 to 8 hours) when uphill. The charge suffices for 20 hours, day duration is 25 hours. Events $a,b,e,f$ happen at a high speed and lead to obesity.    
\label{fig:robot}}
\end{figure} 

The paper is structured as follows. In \cref{sec:back}, we recall some useful notions on timed automata and introduce an operation of squeezing on regular languages. In \cref{sec:band}, we recall our definition of bandwidth and state the problem of bandwidth computation for timed regular languages. 
In main \cref{sec:main}, we compute the bandwidth of obese timed languages. We conclude and briefly discuss the complexity aspects and  perspectives in \cref{sec:con}. 

For the sake of clarity, we provide all the details of the algorithm in the body of the article but postpone 
most of the proofs to the Appendix.
 
\section{Background on timed and finite automata}\label{sec:back}
   
$\pset{X}$ denotes the powerset of $X$ and $\Pset{X}$ removes the empty set from $\pset{X}$.
We employ words, languages, and automata on both alphabets $\pset{\Sigma}$ and $\Sigma$. 
Each word $w=a_1\dots a_n\in \Sigma^*$ can be also seen as $\{a_1\}\dots \{a_n\}\in \pset{\Sigma}^*$. 
Abusing set notation, we omit commas between braces, e.g., shortening $\{a,b\}\{b,c\}$ to $\{ab\}\{bc\}$.
We denote the set of all letters in $w$ by $\lett{w}\in \pset{\Sigma}$. Finally, $\Log\subseteq \real$ stands for the set of logarithms of algebraic numbers. 

%\subsubsection{Timed words, languages and automata}
%\emph{Timed automata} (TA) have been introduced in \cite{AD} for modeling and verification of real-time systems.  
\begin{definition}
    Given $\Sigma$, a finite alphabet of discrete events, a \emph{timed word} over $\Sigma$ is an element from $\left(\Sigma\times\real_+\right)^*$ of the form $w = (a_1,t_1)\dots (a_n,t_n)$, 
with $0 \leq t_1 \leq t_2 \cdots \leq t_n$. We denote $\length(w) = t_n$.
A \emph{timed language} over $\Sigma$ is a set of timed words over the same alphabet. Given a timed language $L$ and $T \in [0,\infty)$, the  corresponding \emph{time-restricted language} is
$L_T = \{ w \in L \mid \length(w)  \leq T\}$.
\end{definition}
% Let $\D$ be such an SCC. Let $Y$ be the set of clocks that are not reset along $\D$. First, remark that there is at least one such clock, the heartbeat clock $h$ (for resetting this clock, at least one black transition has to be taken). Because the automaton is in region-split form, for every location $q$, the region $S(q)$ satisfies $d_y(q)<y<d_y(q)+1$ or $d_y(q)=y$ for any $y\in Y$. Since the cl

For a set of variables $X$, let $G_X$ be the set of finite conjunctions of constraints of the form $x \sim b$ and $x \sim y+b$ with $x,y \in X$, $\sim \in \{<,\leq, >, \geq\}$ and $b \in \nat$.
We rely here on the following form of timed automata:
\begin{definition}[\cite{AD}, variant from \cite{3classes}]\label{def:TA}
 A  TA %(without \linebreak $\varepsilon$-transitions) 
 is a tuple 
$(Q, X, \Sigma, \Delta, S, I, F)$, with
$Q$  the finite set of \emph{locations},
$X$ the finite set of \emph{clocks},
$\Sigma$  a finite \emph{alphabet},
$S,I,F: Q \rightarrow G_{X}$   resp.~the \emph{starting}, \emph{initial}, and \emph{final} clock constraints,
and 
the \emph{transition relation} (set of \emph{edges})\; $\Delta \subseteq Q \times Q \times \Sigma \times G_{X} \times 2^X$.
%
% A timed automaton is \emph{deterministic}  if $\{ (q,\x) : \x\models I(q)\}$ is a singleton and for any two edges 
% $(q,q_1,a,\guard_1,\reset_1)$ and $(q,q_2,a,\guard_2,\reset_2)$ with $q_1 \neq q_2$, the constraint $\guard_1 \wedge \guard_2$ is   non-satisfiable. 
\end{definition}
A TA is \emph{bounded} whenever there is a constant $M$ such that all the starting conditions $S(q)$ 
and guards $\guard_\delta$ require all clock values to be smaller than $M$.

%\paragraph{Semantic details and notations}

The semantics of a TA is a \emph{timed transition system} whose states are
tuples $(q,\x)$ composed of a location $q\in Q$ and a clock valuation (vector) $\x \in [0,\infty)^X$.
We denote $\x[\reset]$ the operation of resetting 
the clocks in $\reset \subseteq X$.
Each edge $\delta  = (q,q', a,\guard, \reset) \in \Delta$ generates timed transitions $(q,\x) \trans{\delta,d} (q', \x') $ where 
$ \x \models S(q)$, $\x+d \models \guard$, $\x' = (\x+d)[\reset]$ and $\x' \models S(q')$.
%$x'_c=0$ whenever $c\in\reset$ and $x'_c=x_c+d$ otherwise, provided that $\x' \models S(q')$. 
Here, $\x+d = (x_1+d,\dots, x_n+d)$.
%\bernardo{L'opération est utilisée juste avant d'être définie}

\emph{Paths} are sequences of edges that agree on intermediary locations  
%At the semantic level, they generate 
and \emph{runs} are sequences of timed transitions that agree on intermediary states.
An \emph{accepting run} is a run $\rho = (q_0,\x_0) \trans{(\delta_1,d_1)} (q_1,\x_1)\cdots \trans{(\delta_n, d_n)}(q_n,\x_n)$,
in which 
%the first state satisifes 
$\x_0 \models I(q_0)$ and 
%the last state satisfies 
$\x_n \models F(q_n)$.
Given a run $\rho$ as above, we define
$\Path(\rho)\triangleq\delta_1\dots \delta_n$. 
Furthermore, if $\delta_i = (q_i,q_{i+1},a_i,\guard_i,\reset_i)$, then 
the timed word associated with $\rho$ is defined as $\Word(\rho)\triangleq %(a_0,0)
(a_1,d_1)(a_2,d_1+d_2)\ldots  (a_n,\sum_{i\leq n}d_i)$. %\bernardo{La lettre $(a_0,0)$ n'est pas définie, ne faut-il pas l'enlever ?}
The \emph{duration} of $\rho$ is $\length(\rho) = \sum d_i$.
The language  $L(\aut)$ of a TA $\aut$ is the set of timed words associated with some accepting run.

% For a small example, consider a 3-edge path $q\to p\to q$ in the third timed automaton in \cref{fig:3kinds}. One of its runs is 
% $$(q,0,0) \trans{(a,0.8)} (p,0,0.8) \trans{(b, 0.7)}(q,0.7,0)\trans{(b,0.2)} (p,0,0.2),$$
% %(q,0,0) \trans{(\delta_1,0.8)} (p,0,0.8) \trans(\delta_2, 1.5)}(q,0.7,0)\trans{(\delta_1,1.7)} (p,0,0.2),
% and the corresponding timed word is $(a,0.8) (b,1.5) (b,1.7)$.

%Whenever we are not interested in timed languages but only in the reachable states, we consider timed graphs
% A \emph{timed graph} is a timed automaton with a singleton alphabet. In a timed graph, we are only interested 
% in the reachable states. 

%their paths and runs are defined similarly to TA:
% \begin{definition}
 A \emph{timed graph} is  a tuple $(Q, X, \Delta, S)$ with $Q,X,S$ as in \cref{def:TA} and $\Delta \subseteq Q \times Q  \times G_{X} \times 2^X$. 
 It is \emph{time-divergent} whenever the number of edges in runs of duration $\leq  T$ is bounded for every $T$. 
 %
 %It is \emph{bounded} whenever all formulas from $G_X$ employed on transitions utilize integers $\leq M$ for some $M \in \nat$.
% all the starting conditions $S(q)$ and guards $\guard_\delta$ require all clock values to be smaller than $M$.%\todo{addresses reviewer 3's comment $\to$ ok?} YES
% \end{definition}

%\subsubsection{Regions and corner-points}

% Many algorithms related to timed automata utilize the finitary abstraction of the timed transition system called the \emph{region construction} \cite{AD} that we briefly recall here.
% and, for a clock vector $\x\in [0,\infty)^X$, $x_{\Int}\triangleq\{c\in X:x_c\in[0,M]\cap\nat\}$, and $x_{\Frac}\triangleq\{c\in X: x_c \in [0,M]\setminus\nat\}$.
%\begin{definition}[Alur \& Dill]
 Clock vectors $\x$ and $\y$ are \emph{region-equivalent} \cite{AD} if 
%\begin{definition}[\!\cite{AD}]
%For a clock vector $\x$, we define $x_\infty=\{c\in X:x_c>M\}$ and $x_\nat=\{c\in X: x_c \in \nat\}$.
%
%Two clock vectors $\x,\y\in\real_+^X$ are \emph{region-equivalent} whenever
 %itemizeetoile
%\begin{itemize*}
   % \item $x_{\Int}=y_{\Int}$ and $x_{\Frac} = y_{\Frac}$;
    %\item 
    for any clock $ c,\  \lfloor x_c \rfloor = \lfloor y_c \rfloor$ and $\{x_c\} =0$ iff  $\{y_c\} =0$;
    %\item 
     and for any two clocks  $c_1, c_2$, $\{x_{c_1}\}\leq \{x_{c_2}\}$ iff $\{y_{c_1}\} \leq \{y_{c_2}\}$ (where $\{x\}$ denotes the fractional part of $x$). 
 %   \end{itemize*}
%\emph{Regions} are equivalence classes of the region equivalence described just above.
%\end{definition}
%
Equivalence classes w.r.t. region-equivalence (called \emph{regions})
%(where any vector $x$ satisfies $x_\Int\cup x_\Frac=X$) 
are simplices of dimension $d\leq \#X$.% and the following definitions coming from \cite{alive} can be stated about its vertices.

% We recall that a directed graph is \emph{strongly connected} whenever each vertex is reachable from each other following edges. Maximal strongly connected subgraphs are called \emph{strongly connected components} (SCCs). An SCC is \emph{non-trivial} whenever it contains a cycle. 

\subsubsection{Optimal reward-to-time ratio in finite and timed graphs.}\label{sec:alive}
Consider a finite directed graph where each edge $e$ is associated with a \emph{reward} $w(e)$ and \emph{time} $t(e)$. 
For a path $\pi$, its reward $w(\pi)$ is the sum of rewards of its edges, similarly for time.  A classical optimization problem is maximizing the reward-to-time ratio $w(\pi)/t(\pi)$ over long or infinite paths. The maximal asymptotic ratio can be attained by iterating some optimal simple cycle. We refer the reader to \cite{ratio} for a survey of exact and approximation algorithms for finding the optimal cycle and/or computing the best ratio.

% A remark on algorithmic aspects is in order. We will consider graphs with integer times and weights in $\Log$. %Due to \cref{prop:log} 
% Hence, the cycle ratios are also in $\Log$. It follows that the brute force algorithm enumerating all the simple cycles and comparing their ratios
% %\todo{Catalin : cherche des références de complexité de comparaison des nombres algébriques}
% would find the exact value of the maximal ratio as a member of $\Log$.

%  Alternatively, the computation of a numeric solution with $k$ decimals would be polynomial in $k$ and the size of the graph, using one of the classical algorithms.
% %\eugene{dire quelque-chose concret sur  la complexité}

% \begin{lemma}[see \cite{ratio}]
%   Given a finite directed graph  with non-negative rewards and  times associated to edges, and sum of times $\geq 1$ for every cycle, let $\alpha>0$ be the maximal reward-per-time ratio over all simple cycles. Then  for all paths $\pi$, 
% $
% w(\pi)\lessapprox\alpha t(\pi).
% $
% \end{lemma}
% %\eugene{check}

The authors of \cite{alive} have extended the  theory of reward-to-time ratio  to timed graphs.  Let us phrase their result in a form suitable for our investigations. %(in particular, we are maximizing the ratio while they were minimizing). 
A \emph{weighted timed graph} (WTG) is just a timed graph together with a reward function $w:Q\to\real_+$.
The reward of a transition $p\trans{\delta,t}q$ (or rather of a stay of $t$ time units in $p$) is $w(p)t$; the reward $w(\rho)$ of a run $\rho$ is the sum of rewards of all its transitions, its  ratio is $w(\rho)/\length(\rho)$. 
\begin{lemma}[\cite{alive}, variant]\label{lem:alive}
Given a bounded  time-divergent WTG, one can compute the optimal ratio $\alpha\geq 0$ and a constant $C>0$ such that 
for all runs $\rho$,
$
w(\rho)\leq \alpha \length(\rho)+C;
$ and
 for any  $T>0$ there exists a run  $\rho$ with $\length(\rho)\leq T$ and $w(\rho)>\alpha T -C$.

\end{lemma}
%The algorithm in \cite{alive} has been stated for integer  weights, but we need it for weights in $\Log$. 
Careful analysis of the algorithm in \cite{alive} 
shows that, when all $w(e),t(e) \in \Log$,  
the resulting  $\alpha$ belongs to $\Log$, 
and can be computed both using some representation of the elements of $\Log$ 
or as a real number with arbitrary  precision.

% The former follows from the proof of \cite[Prop.~4]{alive},  the latter from  
% \cite[Prop.~5]{alive}.  

\subsubsection{Growth rate of regular languages and squeezing.}\label{sec:squeeze}
For a language $L$ on alphabet $\Sigma$ its (logarithmic)  \emph{growth rate} is defined as follows \cite{ChomskyMiller}:
$$
\gro{L}=\limsup_{n\to \infty}\left({\log\#L_n}\right)/{n} \hspace*{20pt} \text{ (where $L_n= L\cap \Sigma^n$).} 
$$
%In \cite{ChomskyMiller}, the growth rate has been characterized for regular languages. 
For a finite automaton $\aut=(Q,\Sigma,\Delta,I,F)$, with $Q=\{q_1,\dots,q_n\}$, we define its $n\times n$ \emph{ adjacency matrix} $M_\aut=(m_{ij})$ with $m_{ij}=\# \{a\in\Sigma:(q_i,a,q_j)\in\Delta\}$.
\begin{thm}[\cite{ChomskyMiller}]\label{thm:chomsky}
Let $\aut$ be a trim\footnote{with all the states reachable from an initial and co-reachable to a final one} deterministic finite  automaton. Then\linebreak $\gro{L(\aut)}$ 
equals the logarithm of the spectral radius\footnote{the max of the eigenvalue moduli}  of  $M_\aut$. 
\end{thm}

% \subsubsection{Squeezing regular languages.}
% \label{sec:squeeze}
A word $W$ over an alphabet $\pset\Sigma$  can be \emph{squeezed}, yielding a word over $\pset \Sigma$:
\begin{itemize}
    \item first factorize it arbitrarily: $W=W_1W_2\dots W_k$ where $W_i\in \pset\Sigma^*$;
    \item then transform each factor $W_i$ into the union of its letters (as subsets of $\Sigma)$.% $\bigcup {W_i}$. 
\end{itemize}

Given a language $L\subseteq \pset{\Sigma}^*$, we define its squeezing $\squeeze L\subseteq \pset{\Sigma}^*$, a language consisting of all squeezings of all the words in $L$. We also use a notation for its $n$-letter fragment defining $\squeeze_n L=\squeeze L\cap \pset{\Sigma}^n$.
Hence,  $\{abc\}\{b\}\emptyset\{abc\}\{a\} \in \squeeze \big(\{ab\}\{ac\}\{b\}\{c\}\{abc\}\emptyset\{a\}\{a\}\big)$.
%$\{ab\}\{ac\}\{b\}\{c\}\{abc\}\emptyset\{a\}\{a\}$ can be squeezed as 
%$\{ab\}\{ac\}|\{b\}||\{c\}\{abc\}|\emptyset\{a\}\{a\} \to \{abc\}\{b\}\emptyset\{abc\}\{a\}$.
We abuse notation and apply squeezing also to languages over $\Sigma$, hence $\{ab\}\{abc\}\emptyset\{a\}\{ab\} \in \squeeze(abacccbacbaba)$.

%Squeezing of a word over $\Sigma$ also yields a word over $\pset \Sigma$, e.g.~$abacccbacbaba$ can be squeezed as
%$ab|acccb| |acb|a|ba \to \{ab\}\{abc\}\emptyset\{a\}\{ab\}$.

\begin{figure}[t]\scriptsize
\begin{tikzpicture}
\node[state](px)[initial]  at (0,4)  {$p$};
\node[state](qx)[accepting] at (1.5,4) {$q$};
\draw [post] (px) edge [bend left,above] node {$a,b$} (qx);
\draw [post] (qx) edge [below] node {$c$} (px);
\draw  (qx) edge [loop above] node {$b,c$} (qx);
\end{tikzpicture}
\begin{tikzpicture}
\node[state](p)[initial]  at (0,1)  {$p$};
\node[state](q)[accepting] at (2.4,1) {$q$};
%\node[state](r) at (2,0) {$r$};
\draw  (p) edge [loop above] node {$\emptyset,ac,bc,abc$} (p);
\draw  (q) edge [loop above] node {$\emptyset,b,c,bc,ac,abc$} (q);
%\draw  (r) edge [loop left] node {$\emptyset$} (r);
\draw [post,bend left] (p) edge [above] node {$a,b,ac$}  node[below] {$ab,bc, abc$} (q);
\draw [post,bend left]  (q) edge [above] node {$c,bc,ac,abc$} (p);
\end{tikzpicture}
\begin{tikzpicture}
\node[state](p)[initial]  at (0,1)  {$p$};
\node[state](q)[accepting] at (3.4,1) {$q$};
\node[state](pq)[accepting] at (1.7,2) {$pq$};
\draw  (p) edge [loop above] node {$\emptyset$} (p);
\draw  (q) edge [in=30,out=60,loop] node[above] {$\emptyset,b$} (q);
\draw  (pq) edge [loop above] node {$\emptyset,c,ac,bc,abc$} (pq);
\draw[post] (p) edge[above] node {$a,b,ab$} (q);
\draw [post] (p) edge[above,sloped] node {$ac,bc,abc$} (pq);
\draw [post](q) edge[above,bend right=1.5cm,sloped] node {$c,bc,ac,abc$}(pq);
\draw [post](pq) edge[above,sloped] node {$a,b,ab$} (q);
\end{tikzpicture}
\caption{\small (Left) an automaton for $(a+b)(b+c+ca+cb)^*$; (middle) an automaton for its squeezing (sets are written without braces); (right) its determinization.  \label{fig:squeeze}
}
\end{figure}
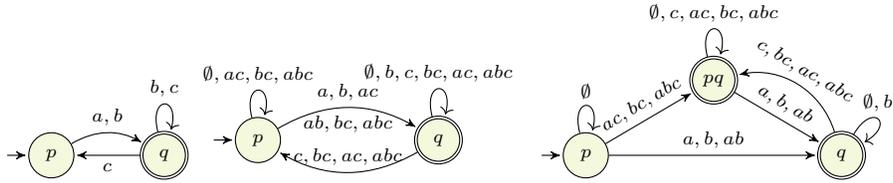
%\begin{proposition}\label{prop:squeeze}
If $L$ is regular, then $\squeeze L$ is regular too.
%\end{proposition}
The following construction, illustrated in \cref{fig:squeeze}, yields an automaton recognizing  $\squeeze L$ from one recognizing  $L$:

\begin{con} \label{con:squeeze}
Given  $\aut=(Q, \pset{\Sigma},\Delta,I,F,)$, $\squeeze L(\aut)$ is accepted by $\squeeze \aut=(Q,\pset{\Sigma},\Delta',I,F)$, 
where $(p,B,q)\in \Delta'$ whenever $\aut$ has a path from $p$ to $q$ with the union of its labels equal to $B$. 
In particular, $\squeeze \aut$ has self-loops labeled by $\emptyset$ on each state. 
\end{con}
%
%Combining squeezing, determinization, and computation of the spectral radius, we obtain the following result. %\cref{thm:chomsky,prop:squeeze}.
\begin{proposition}
Given a finite automaton $\aut$, the growth rate of its squeezed language  $\gro{\squeeze L(\aut)}$ belongs to $\Log$ and is  computable from $\aut$.
\end{proposition}
So, for the automaton $\aut$ on the left of \cref{fig:squeeze} 
(used later for bandwidth computation), 
the growth rate of $\squeeze L(\aut)$ is the logarithm of the spectral radius of the adjacency matrix of the automaton on the right:
$$
\gro{\squeeze L(\aut)}= \log \rad \begin{pmatrix}
1 & 3 & 3\\
0 & 2 & 4\\
0 & 3 & 5
\end{pmatrix}=\log((7+\sqrt{57})/2)\approx 2.863.
$$

\section{Bandwidth of timed languages and the main problem}
\label{sec:band}
 \subsubsection{Pseudo-distance on timed words.}
The notion of bandwidth of timed languages relies on a pseudo-distance on timed words defined in \cite{distance}.
Here we extend it to $\pset{\Sigma}$ as follows
(where $\min \emptyset = \infty$):

%the pseudo-distance on timed words introduced in \cite{distance}.
\begin{definition}
Given 
$w = (A_1,t_1)\dots(A_n,t_n)$ and $v = (B_1,s_1)\dots(B_m,s_m)$ 
two timed words over $\pset{\Sigma}$,
the \emph{pseudo-distance} $d(w,v)$   is defined as follows:
\[
 \dr(w,v)\triangleq \max_{\substack{i\in\{1..n\} \\ a \in A_i}}
 \min_{\substack{j\in\{1..m\}\\ b \in B_j}} \{ |t_i-s_j|:a=b \};\quad 
 d(w,v) \triangleq  \max( \dr(w,v) , \dr(v,w) ). 
\]
\end{definition}
%In the basic case of words on $\Sigma$,
%$w = (a_1,t_1)\dots(a_n,t_n)$ and $v = (b_1,s_1)\dots(b_m,s_m)$ this yields 
%Clearly, when $A_i = \{a_i\}$ and $B_i = \{b_i\}$, we get the pseudo-distance from \cite{distance}.

% $$
%  \dr(w,v)\triangleq \max_{i\in\{1..n\}}\min_{j\in\{1..m\}} \{ |t_i-s_j|:a_i=b_j \}; \quad
%  d(w,v) \triangleq  \max( \dr(w,v) , \dr(v,w) ).
% $$

\begin{figure}[t]
\begin{center}
{
\usetikzlibrary {arrows.meta,positioning} 
\begin{tikzpicture}
\draw (0,.8) -- (.7,.8) \aaa  --(1.8,.8) \bbb  -- (3,.8) \aaa-- (4,.8) \bbb-- (4.7,.8) \aaa --
(5,.8)node[anchor=west]{$\scriptstyle{u=(\{abc\},0.7),(\{ab\},1.8), (\{bc\},3), (\{a\},4), (\{ab\},4.7)}$};

\draw (0,0) -- (.6,0) \aaa   -- (1,0) \aaa--(1.7,0) \bbb  -- (3,0) \aaa-- (4.1,0) \aaa-- (4.6,0)\bbb --
(5,0)node[anchor=west]{$\scriptstyle{v=(\{ab\},0.6),(\{bc\},1), (\{ac\},1.7), (\{b\},3), (\{b\},4.1),(\{a\},4.6)}$};

\draw [dotted]
%, arrows={ ->[width=1mm,length=1.2mm,sep=1.8mm]}]
(.7,.8) edge[left] node{$\scriptscriptstyle a,b$} (.6,0)
(.7,.8) edge[above] node{$\scriptscriptstyle c$} (1,0)
(1.8,.8) edge[below] node{$\scriptscriptstyle b$} (1,0)
(1.8,.8) edge[right] node{$\scriptscriptstyle a$} (1.7,0)
(3,.8) edge[right] node{$\scriptscriptstyle a$} (1.7,0)
(3,.8) edge[left] node{$\scriptscriptstyle b$} (3,0)
(4,.8) edge[left] node{$\scriptscriptstyle a$} (4.1,0)
(4.7,.8) edge[above] node{$\scriptscriptstyle b$} (4.1,0)
(4.7,.8) edge[right] node{$\scriptscriptstyle a$} (4.6,0)
;

%ancienne version fausse
% \draw [dotted, arrows={
% ->[width=1mm,length=1.2mm,sep=1.8mm]}]
% (.6,0) edge (.7,.8)
% (.7,.8) edge (.6,0)
% (1,0) edge (.7,.8)
% (1.7,0) edge (1.8,.8)
% (1.8,.8) edge (1.7,0)
% (3,0) edge (3,.8)
% (4,.8) edge (4.2,0)
% (3,.8) edge (3,0)
% (4.2,0) edge (4,.8)
% (4.1,0) edge (4.1,.8)
% (4.1,.8) edge (4.1,0)
% ;

\end{tikzpicture}
}
\end{center}
\caption{\small Pseudo-distance between two timed words over $2^\Sigma$, 
$\protect\dr(u,v)=0.2$. Dotted edges represent the closest position for matching letter.
%\  \protect\dr(v,u)=0.3$, thus  %
%$d(u,v)=0.3$.
}\label{fig:dist}
\end{figure}
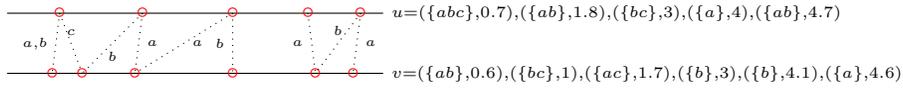

Computing $d$ is illustrated in \cref{fig:dist}. Intuitively, 
an observer who tries to distinguish timed words $u$ and $v$
will pair each  timing of an event in $u$ to the closest timing in $v$ ``emitting'' the same event, 
and deduce that $u$ and $v$ are close when all these pairs of timings are close enough. 
We symmetrize this distance, similarly to  \cite{distance}.
Also, 
% similar to the original definition, 
$d$ may fail to distinguish timed words, that is $d(w_1,w_2)=0$  but $w_1\neq w_2$, which is the reason why $d$ is only a pseudo-distance.
%this could happen when $w_1$ and $w_2$
%only differ by order and quantity of simultaneous letters, e.g.~for $w=(a,1)(b,1)$ and $v=(b,1)(b,1)(a,1)$, that is why $d$ is  only a pseudo-distance.

%This pseudo-distance allows a meaningful comparison of timed words with a different number of events. It is illustrated in \cref{fig:dist}.
%Intuitively, two words are close to each other when they cannot be distinguished by an observer that reads the discrete letters of the word exactly 
%(they can determine whether or not a letter has occurred) but with some imprecision w.r.t.~time. 
%So, the observer cannot determine when two letters are very close to one another, which one came before the other, and not even how many times a letter was repeated within a short interval. 

\subsubsection{$\varepsilon$-capacity and $\varepsilon$-entropy.}
These two closely related notions from  \cite{kolmoEpsilon} characterize the quantity of information needed to describe  (with precision $\varepsilon$) any element of a given space. 
We first recall a couple of definitions (adapted to pseudo-distances).
\begin{definition}\label{def:sep:net}
Let $(X,d)$ be a (pseudo-)metric space, a subspace of some space $Y$. A subset $M\subseteq X$ is called $\varepsilon$-\emph{separated} whenever each $x, y\in M$ with $x\neq y$ satisfy $d(x,y)>\varepsilon$. 
A set $N\subseteq Y$ is an $\varepsilon$-\emph{net} for $X$ (in $Y$) whenever for all $x\in X$ there exists $y\in N$ with $ d(x,y)\leq \varepsilon$.
\end{definition}
For a totally bounded space \cite{kelley2017general} $X$ and any $\varepsilon >0$, a finite $\varepsilon$-net must exist within $X$ itself, and cardinalities of $\varepsilon$-separated sets are bounded, which justifies:% the following.% definition.
\begin{definition}[\cite{kolmoEpsilon}]\label{def:ent:cap}
Given a totally bounded (pseudo-)metric space $X\subseteq Y$ let $\mathcal{M}_\varepsilon(X)$ be the maximal size of an \emph{$\varepsilon$-separated} subset of $X$. Then the \emph{$\varepsilon$-capacity} of $X$ is  defined as 
$
\capa_\varepsilon(X)= \log \mathcal{M}_\varepsilon(X).
$
Let now $\mathcal{N}_\varepsilon^Y(X)$ be the minimal size of an $\varepsilon$-net for $X$ (in $Y$). Then the $\varepsilon$-entropy of $X$ (in $Y$) is 
$
\ent^Y_\varepsilon(X)= \log \mathcal{N}_\varepsilon^Y(X).
$
\end{definition}

As shown in \cite{kolmoEpsilon},  $\varepsilon$-entropy and $\varepsilon$-capacity are related by inequalities
\begin{equation}\label{prop:inequality}
\ent^Y_\varepsilon(X)\leq \capa_\varepsilon(X)\leq \ent^Y_{\varepsilon/2}(X).    
\end{equation}

\subsubsection{Problem statement. }
Let us introduce the central notion for this article:
%We recall the definition from \cite{bw-our}, central for this article, as it was for \cite{3classes,CIAA}.
\begin{definition}[\cite{bw-our}] The $\varepsilon$-\emph{entropic bandwidth} and $\varepsilon$-\emph{capacitive bandwidth} of a timed language $L$ are defined respectively as $\bandh_\varepsilon(L)=\limsup_{T\to\infty} \ent^U_\varepsilon(L_T)/T$  and $\bandc_\varepsilon(L)=\limsup_{T\to\infty} \capa_\varepsilon(L_T)/T.$
\end{definition}

Here $U$ stands for the universal timed language. Remark that $(L_T,d)$ is totally bounded (\cite{distance}), hence $\ent^U_\varepsilon(L_T)$ and $\capa_\varepsilon(L_T)$ are well defined.
%
%Here, $\varepsilon$-nets (and thus  $\varepsilon$-entropy) are considered within the universal timed language $U$. 
A couple of simple properties should be mentioned: it follows from \eqref{prop:inequality} that
$
\bandh_\varepsilon(L)\leq \bandc_\varepsilon(L)\leq \bandh_{\varepsilon/2}(L)$;
from \cite[Thm.~2]{distance} follows the upper bound valid for any timed language $L$ (over a fixed alphabet $\Sigma$): $\bandc_\varepsilon(L)= O({1}/{\varepsilon})$.

%  there is an upper bound on the bandwidth of timed languages:
%  \begin{proposition}[\protect{\cite[Thm.~2]{distance}}]\label{prop:max}
%  The bandwidth of the universal timed language $U_\Sigma$ satisfies $\bandc_\varepsilon(U_\Sigma)=\Theta(\frac{1}{\varepsilon})$.  Hence, for any timed language $L$, the upper bound holds: $\bandc_\varepsilon(L)= O(\frac{1}{\varepsilon})$.
% \end{proposition}

%\begin{MP}
%Given a timed automaton $\aut$, compute the $\varepsilon$-bandwidths of its language asymptotically, as $\varepsilon\to 0$.
%\end{MP}

We  have shown in \cite{3classes} that there are three classes of timed regular languages: meager with bandwidth $O(1)$, 
normal with bandwidth $\Theta(\log 1/\varepsilon)$, and obese with the maximal possible bandwidth $\Theta(1/\varepsilon)$. 
In the rest of the paper, we deal with a particular case of \eqref{challenge}: the computation of the bandwidths of obese TA.
\begin{MP}
Given an obese TA $\aut$, compute a real number $\alpha$ 
such that %$\bandc_\varepsilon(L(\aut))\eqapprox \alpha/\varepsilon$, i.e. 
$\bandc_\varepsilon(L(\aut)) = (1+o(1))\alpha/\varepsilon$ for small $\varepsilon$. 
\end{MP}
We are equally interested in the asymptotic behavior of $\bandh$.
%\catalin{promu en pb principal, mais il faudrait éliminer l'autre main problem duquel on ne s'occupe pas du tout !}

%
\section{Bandwidth computation for obese automata}\label{sec:main}

\subsection{Trivially timed languages}
% A timed automaton is called \emph{trivially timed} if all its guards and starting conditions are $\mathbf{true}$. Such automata yield the simplest examples of obese languages. 
% %\catalin{vont servir de briques Lego pour le reste} 
% \begin{con}\label{con:supp}
%  For a timed automaton $\aut=(Q, X, \Sigma, \Delta, S)$, we define its \emph{support} as a finite automaton $\supp \aut=(Q,  \Sigma, \Delta')$, where for each  edge $(q,q', a,\guard, \reset) \in \Delta$ we put $(q,q', a)$ in $\Delta'$. 
%  For a timed automaton $\aut$, we obtain its trivially timed version $\widetilde\aut$ by replacing all guards and starting conditions by $\mathbf{true}$.
% \end{con}

A TA is called \emph{trivially timed} if all its guards and starting conditions are $\mathbf{true}$. Such automata yield the simplest examples of obese languages. 

For a TA $\aut=(Q, X, \Sigma, \Delta, S)$, we define its \emph{support} as a finite automaton $\supp \aut=(Q,  \Sigma, \Delta')$, where for each  edge $(q,q', a,\guard, \reset) \in \Delta$ we put $(q,q', a)$ in $\Delta'$. 
For any TA $\aut$, we call the \emph{trivially timed version} of it, the automaton $\widetilde\aut$ obtained by replacing all guards and starting conditions by $\mathbf{true}$.%, while keeping the same support.

The following notation will simplify our statements on asymptotic behaviors up to a multiplicative or additive constant, where $\varepsilon\to 0$ and uniformity w.r.t.~ all other parameters is required. 
\begin{definition}[Asymptotic notation]
Given  two positive-valued  functions $f(\varepsilon,x)$ and $g(\varepsilon, x)$ (with any kind of arguments $x$), we write $f\lesssim g$ whenever 
\[
\forall \varkappa>0\; \exists c,\varepsilon_0>0,\forall\varepsilon<\varepsilon_0,\forall x: \, f(\varepsilon,x)< c  g^{1+\varkappa}(\varepsilon,x).
\]
%
% for all $\varkappa>0$ exist $c$ and $\varepsilon_0>0$ such that for all $\varepsilon<\varepsilon_0$ and all $x$ it holds that $f(\varepsilon,x)< c  g^{1+\varkappa}(\varepsilon,x)$.  
%
For real-valued  $f(\varepsilon,x)$ and $g(\varepsilon, x)$, we write $f\lessapprox g$ whenever 
\[
\forall \varkappa>0\; \exists c,\varepsilon_0>0,\forall\varepsilon<\varepsilon_0,\forall x: \, f(\varepsilon,x)< c+ (1+\varkappa)g(\varepsilon,x).
\]
We write $f\eqsim g$ if $f\lesssim g$ and $g\lesssim f$, and $f\eqapprox g$ if $f\lessapprox g$ and $g\lessapprox f$.
\end{definition}

%Hence $f\lessapprox g$ iff $f\leq g +o(g)+O(1)$ as $\varepsilon\to 0$ and uniformly in other parameters.

We can now characterize the entropy and capacity of trivially timed languages, which will play a key role for the general case. 
\begin{restatable}{proposition}{proptrivial}\label{prop:trivial} 
Let $\aut$ be trivially timed with $\supp\aut$ strongly connected. 
%and  the growth rate of its squeezed language  
Denote $\alpha = \gro\squeeze (\supp \aut)$. 
%Let $L$ be the timed language of $\aut$ w.r.t.~some nonempty sets of initial and final locations. 
Then entropy and capacity of $L_t(\aut)$ satisfy $\ent^U_{\varepsilon/2}(L_t) \eqapprox  \capa_{\varepsilon} (L_t) \eqapprox  {\alpha t}/\varepsilon$ for $t>0$.
Also, $\alpha$ is independent of the choice of initial or final locations. 
\end{restatable}
As a consequence, $L$ is obese, the  bandwidth satisfies $\bandh_{\varepsilon/2}(L) \eqapprox  \bandc_{\varepsilon} (L)\eqapprox  \alpha / \varepsilon$.

\subsection{Bandwidth-preserving preprocessing of timed automata}
Given a TA (possibly obese), we will first proceed with a couple of bandwidth-preserving transformations, as well as the classification of its transitions into red ones and black ones. Informally, high bandwidth is produced within red SCCs (``spots''), and black transitions are used to connect them and to loop.

%\subsubsection{Adding heartbeat} to contain high frequency runs within 1 time unit intervals.

%\subsubsection{Adding heartbeat and urgency clock} We will slightly modify the timed automaton and ``beat out the rhythm'' by emitting a special event once each time unit.

\subsubsection{Adding heartbeat and urgency clock.}
\begin{con}
Given a TA, we add two features:
\begin{description}
\item[Heartbeat:] a new clock $h$  and a new letter $b$; next,  we add to each transition the constraint $h\leq 1$, and  to each location $p$ a self-loop $p\trans{b,h=1,\{h\}} p$.
\item[Urgency clock:] a new  clock $u$, reset  at every transition, never tested in guards.
\end{description}
\end{con}
A heartbeat is emitted every time unit and contains  high-frequency runs within 1 time unit intervals. At the next stage of the algorithm, the urgency clock will help discriminating urgent transitions. The two features do not change the bandwidth; the latter even  preserves the language.

% Quite naturally forced beat does not add information to the language:
% \begin{restatable}{lemma}{lemheartband}\label{lem:heart:band}
% Adding the heartbeat does not change the bandwidth.
% \end{restatable}
% %
% \subsubsection{Adding urgency clock} to help discriminating urgent transitions.
% %
% This transformation is even simpler and it does not change the timed language.%but serves at the next two stages. 
% \begin{con}
% Given a TA, we add a new \emph{``urgency''} clock $u$, reset it at every transition, and do not test it.
% \end{con}

\subsubsection{Region-splitting and bounding.}
This step is a language-preserving transformation of a TA into a form similar to the region automaton \cite{AD}, but typed as a TA. We give a nondeterministic version of the definition  from \cite{entroJourn,3classes,CIAA}.
\begin{definition}\label{def:rs:bounded}
A \emph{region-split TA} (or \RTA) is a TA\ $(Q,X,\Sigma,\Delta,S,I,F)$, which is bounded and such that, for any location $q\in Q$: 
(1) $S(q)$ defines a non-empty region, called the \emph{starting region} of $q$;
(2) all states in $\{q\}\times S(q)$ are reachable from an initial state and co-reachable to a final one;
(3) for any edge $(q,q',a,\guard,\reset)\in\Delta$,  
$\left(\{S(q)+d\mid d\in \real_+\}\cap \guard\right)[\reset]=S(q')$, where we utilize the $(\cdot) [\reset]$ operator lifted to sets of clock valuations.

% \begin{itemize}[nosep]
% %\item the automaton is deterministic;
% \item $S(q)$ defines a non-empty region, called the \emph{starting region} of $q$;% in $[0,M]^X$;
% \item all states in $\{q\}\times S(q)$ are reachable from an initial state and co-reachable to a final one;% state;
% %\item either $I(q)=S(q)$ is a singleton\footnote{by definition of DTA this is possible for a unique location $q$} or $I(q)=\emptyset$;
% \item for any edge $(q,q',a,\guard,\reset)\in\Delta$,  
% $\left(\{S(q)+d\mid d\in \real_+\}\cap \guard\right)[\reset]=S(q')$, where we utilize the $(\cdot) [\reset]$ operator lifted to sets of clock valuations.
% \end{itemize}

%\aldric{$M$ n'intervient pas dans la définition. $\to$ faut-il supprimer la mention de $M$ ou bien (re)-préciser qu'il s'agit de la constante maximale?}
\end{definition}

Any TA having an upper bound on some clock at every transition (a heartbeat $h$ in our case)
%\todo{addresses reviewer 3's comment $\to$ ok? Comment ça s'articule avec le heartbit? $\to$ Ça passe (pas de multiplication infinie des locations)~: pour chaque autre horloge, quand elle est $>M$ dans l'automate d'origine, on est dans la location \emph{bis} du \RTA\ où l'horloge est remise artificiellement à zéro dans toutes les arrêtes sortantes, donc au moins aussi souvent que le heartbeat.} 
can be brought into a bounded region-split form, with the same clocks but an exponentially larger set of locations.
%
% Note that this construction is similar to that of the region automaton introduced in \cite{AD}, except that the outcome is slightly coarser and is typed as a timed automaton.

% A couple of remarks are in order:
% \begin{itemize}
% \item when an automaton with heartbeat is put into a region-split form, then its guards are bounded, as well as starting regions;
%     \item when an automaton with the urgency clock $u$ is put into a region-split form, all  $0$-time transitions (leading to a region with $u=0$) are separated from those taking a positive time and leading to a region with $u>0$.
% \end{itemize}
When a TA with urgency clock $u$ is put into a region-split form, all  $0$-time transitions 
(leading to a region with $u=0$) are separated from those taking a positive time and leading to a region with $u>0$, which makes the next transformation possible.

\subsubsection{Eliminating zeros.} 
 The next transformation removes all urgent transitions.   
\begin{definition}
   In a TA (with urgency clock $u$), a transition $\delta=(p,a,\guard,\reset,q)$ is  \emph{urgent} whenever $\guard\implies (u=0)$.
% \end{definition}
% \begin{definition}
   An \RTA\ is called $0$-\emph{free} when none of its transitions is urgent.
\end{definition}
Given an \RTA\ $\aut$ (with urgency clock) over an alphabet $\Sigma$, it is possible to construct a  $0$-free \RTA\  $\nu\aut$ over $\Pset{\Sigma}$ with the same bandwidth.
\begin{con} [0-elimination, automata] \label{con:0}
 For each path
 $q_0 \trans{\delta_0} q_1\cdots \trans{\delta_k}q_{k+1}$
  with non-urgent $\delta_0$ and urgent $\delta_{1}..\delta_{k}$  such  that $\delta_i=(q_i,q_{i+1},a_i,\guard_i,\reset_i)$,  we add a compound transition
  $
  \delta'=\left(q_0,q_{k+1},\{a_0,\dots,a_n\},\guard_0,\bigcup_{i=0}^k \reset_i\right)
  $.
  Then, we remove the urgent transitions.
  We also make initial all the states reachable from an initial one by urgent transitions only.% Starting and final states remain unchanged. 
\end{con}
% \input{figures/elim}
% The construction is illustrated in \cref{fig:elim}. 
% We state its correctness as follows. 

% We state the correctness of the above construction as follows.

% \begin{proposition} \cref{con:0} is correct, in the sense that, for an \RTA\ $\aut$ with urgency clock, $\nu\aut$ is $0$-free and recognizes $\nu L(\aut)$. 
% \end{proposition}

\subsubsection{The red and the black.}
After preprocessing, we now have a 0-free \RTA\ with the heartbeat and urgency clock, over alphabet $\Pset{\Sigma}$, and we will call it \emph{standard-form timed automaton} (\CTA). Our next aim is to identify its parts producing unbounded frequency information (red) and others (black).

\begin{definition}\label{def:red}
    A cycle in an \CTA\ is \emph{fast} whenever it can be traversed twice in less than $1$ time unit.
    A transition is \emph{red} whenever it belongs to a fast cycle. It is \emph{black} otherwise. 
\end{definition}

Red transitions are illustrated by \cref{fig:avatars}, left.

Only fast cycles may produce symbols with unbounded frequency, and it is immediate from \cite[Thm.~4]{3classes} that  any obese \CTA\ has a fast cycle. % (or equivalently has red transitions). 
Given an automaton with black and red transitions,  we consider the graph of red transitions and  refer to  %nontrivial --- they are always nontrivial
strongly connected components therein as \emph{red SCCs}.

\subsubsection{Splitting the red component.}
%\todo{Improve. Refer to example. $\to$ please review} 
\begin{figure}[t]\scriptsize
%\begin{center}
\begin{tikzpicture}[every node/.style={text=black}]
\node[state](p)  at (0,5)  {$p$};
\node[state](q) at (2.5,5) {$q$};
\node[state](r) at (1.25,0) {$r$};
\draw [post] (p) edge [bend left,above,red] node[near start] {$x,y\in(0,1);\{x\}$} node[below]{$a,b$}  (q);
\draw [post] (q) edge [bend left,below,red] node {$x,y\in(0,1)$}node[above]{$c$} (p);
\draw [post] (q) edge [below,sloped] node {$b;x\in(1,2)$} (r);
\draw [post] (r) edge [below,sloped] node {$b;x\in(6,7);\{xy\}$} (p);
\draw  (q) edge [loop above,red] node {$x,y\in (0,1)$} node[right=1mm]{$b,c$} (q);
\draw  (r) edge [loop below,red] node {$a,b,c;x\in(1,2)$} (r);
\end{tikzpicture}
\hspace{-2mm}
\begin{tikzpicture}[every node/.style={text=black}]
\node[state](px)  at (0,4)  {$p,\{x\}$};
\node[state](qx) at (3,4) {$q,\{x\}$};
%\node[state](p)  at (1,2)  {$p,\emptyset$};
\node[state](q) at (2.7,1.5) {$q,\emptyset$};
\node[state](r) at (1.5,0) {$r,\emptyset$};
\draw [post] (px) edge [bend left,above,red] node {$a,b;\{x\}$} (qx);
\draw [post] (px) edge [below,red,sloped] node {$x,y\in(0,1);\{x\}$}node[above,pos=0.4]{$a,b$\ } (q);
\draw [post] (qx) edge [below,red] node {$c$} (px);
\node at (0.7,5.3) {\large $x,y\in(0,1)$}; 
%\draw [post] (q) edge [below,red] node {$a,x<1$} (p);
\draw [post] (q) edge [right] node {$b;x\in (1,2)$} (r);
%\draw [post] (r) edge [left] node {$b,2<x<3,\{x\}$} (p);
\draw [post] (r) edge [below, sloped] node {$b;x,y\in (6,7);\{x,y\}$} (px);
\draw  (q) edge [loop above,red,text=black] node {$b,c;x,y\in(0,1)$} (q);
\draw  (qx) edge [loop above,red] node {$b,c$} (qx);
\draw  (r) edge [loop below,red] node {$a,b,c;x\in (1,2)$} (r);
\draw[rounded corners,very thick,fill opacity=0.2, fill=pink] (-0.6,3.3) rectangle ++(4.2,2.4); 
\draw[rounded corners,very thick,fill opacity=0.2, fill=pink] (1.65,1) rectangle ++(2.1,1.9); 
\draw[rounded corners,very thick,fill opacity=0.2, fill=pink] (0.3,-1.5) rectangle ++(2.4,2);
\end{tikzpicture}
\hspace{-3mm}
\begin{tikzpicture}[every node/.style={text=black}]
\node[state,rectangle,rounded corners](pxin)  at (0,6.2)  {$\check p:2.863$};
\node[state](px)  at (3,6.2)  {$p$};
\node[state,rectangle,rounded corners](qin) at (1.1,3) {$\check q:2$};
\node[state](q) at (3.15,3) {$q$};
\node[state,rectangle,rounded corners](rin) at (2.3,0) {$\check r:3$};
\node[state](r) at (0.1,0) {$r$};
\draw [post] (px) edge [above,sloped,red] node {$x,y\in (0,1);\{x\}$} (qin);
\draw [post] (q) edge [above,sloped] node {$x\in(1,2)$} (rin);
\draw [post] (r) edge [above, sloped] node {$x\in (6,7);\{xy\}$} (pxin);
\draw[rounded corners,very thick,fill opacity=0.2, fill=pink] (-0.65,5.8) rectangle ++(4,1); 
\draw[rounded corners,very thick,fill opacity=0.2, fill=pink] (0.7,2.5) rectangle ++(2.9,1.1); 
\draw[rounded corners,very thick,fill opacity=0.2, fill=pink] (-0.3,-0.5) rectangle ++(3.1,1);
\draw [post] (pxin) edge [above,thick,blue] node {$x,y\in(0,1);\{x\}$} (px);
\draw [post] (qin) edge [above,thick,blue] node {$x,y\in(0,1)$} (q);
\draw [post] (rin) edge [above,thick,blue] node {$x\in(1,2)$} (r);
\end{tikzpicture}
%\end{center}
\caption{\small Example of bandwidth computation. Left: a timed automaton. Middle: its stratified version, pink rectangles correspond to spots, sink states omitted. Right: its abstraction, blue arrows are abstract transitions.\label{fig:avatars}
}\end{figure}
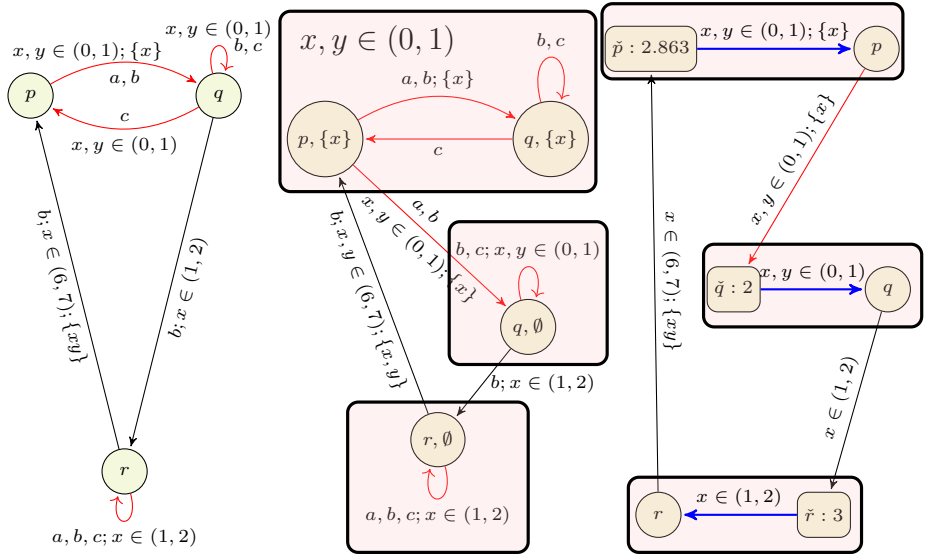
In the next step, we build a \emph{stratified} automaton, in which we materialize different \emph{strategies} for producing bandwidth when staying inside a given red SCC.  
%
%Let us explain the strategies using  the red SCC $\{p,q\}$ in  the example on \cref{fig:avatars}, left.
The example on \cref{fig:avatars}, left, illustrates strategies.
One strategy for the red SCC $\{p,q\}$ consists of wandering with high frequency over all its
% states and 
red transitions (as we will see, this yields an $\varepsilon$-capacity of $\approx 2.863 t/\varepsilon$ in time $t\leq 1$).
%, which cannot exceed $1$ time unit, because the clock $y$ reaches its limit $1$). 
At the end of time $t$, the clock $x$ (recently reset) has  a value close to $0$, and it takes $6$ more time units to travel back to $p$ with the clock $y$ reset.
Another strategy consists of staying in $q$ while emitting with a high frequency events $b$ and $c$. This yields a smaller capacity (only $2 t/\varepsilon$), but the advantage is that $x$ is not reset, and at the end equals $t$. Thus, the travel back to $p$ takes $6-t$ time units, less than in the first strategy. %We also remark that during one stay in the red SCC $\{p,q\}$ it is possible to follow for some time the former strategy, and then switch to the latter (but not the opposite). In fact, this switching corresponds to the last transition from $p$ to $q$ (more generally, the last transition resetting some clock). 
The bandwidth computation should thus consider  and compare all possible strategies in all red SCCs  (and switching between them).

In the following construction of the stratified automaton,  we represent such strategies as subsets $Z$  of the clocks (to be reset, as explained below). For each location $q$ and %each possible strategy, materialized by its 
set of clocks $Z\subseteq X$, we make a location $(q,Z)$ referred to as $Z$-avatar of $q$.
Every $Z$-avatar is committed to resetting all the clocks in $Z$ and no others before taking a black transition.
Switching from a $Z$-avatar to a $Z\setminus \{x\}$-avatar (and switching strategies) is possible when taking a transition that resets $x$. % (the last one in the red SCC, guessed non-deterministically). 
Finally, we leave the red SCC by taking a black transition from some $\emptyset$-avatar.

\begin{con}[stratified automaton]\label{con:augment}
Let $\aut$ be an \CTA.  For a location $q$, let $\resets(q)$ be the set of all the clocks reset in the red SCC containing $q$ (or $\emptyset$ if $q$ is not in a red SCC).  
The  \emph{stratified automaton}  $\aut'$ is constructed as follows:
\begin{itemize}
    \item $Q'$ contains avatars  $(q, Z)$ of each $q\in Q$  for each set of clocks $Z\subseteq \resets(q)$;
    \item for a black transition $p\trans{\delta} q \in \Delta$ we put into $\Delta'$ its avatars
        $(p,\emptyset)\trans{\delta} (q,Z)$ for all  $Z$; %\bernardo{Nécessaire que si p ou q touchent une transition rouge, si on colorie les localisations à l'avance peut-être on peut s'épargner quelques avatars}
    \item  given a red transition $p\trans{\delta} q \in \Delta$, for all sets of clocks  $Z,U$ satisfying $U\subseteq \reset_\delta\subseteq Z$, we put into $\Delta'$ a (red) avatar  of $\delta$, going from  $(p,Z)$ to $(q,Z\setminus U)$;
    \item initial conditions are copied to all avatars: $I'(q,Z)=I(q)$ for all $Z$; similarly for starting conditions $S$;
    \item final conditions are copied in a more restricted way: $F'(q,\emptyset)=F(q)$.
\end{itemize}
 Finally, the automaton is trim: locations unreachable from an initial one and not co-reachable to a final one are removed. 
\end{con}
%
% We remark that an avatar $(q,\emptyset)$ can take a black transition (guessing the target avatar). It cannot take a red one since each of those resets at least the urgency clock $u$.
% %It can also take a red transition, provided this transition does not reset any clock.  
% A location $(q,Z)$ (with $Z\neq\emptyset$) will take a sequence of red transitions, which reset all the clocks in $Z$ but not the others. When a clock is reset the last time in the red sequence, we remove it from the set $Z$. 
% \aldric{C'est partiellement une redite...}

Any \CTA\ and its stratified version have the same languages, see the appendix. 
%
%In a later step, in \cref{sec:abs}, each maximal SCC within which a given strategy applies is abstracted as a single state decorated with a single real number (the bandwidth locally produced by the SCC), yielding a weighted timed graph.

%

\subsubsection{Speedy timed automata.}
%\eugene{est-ce cohérent maintenant?}
We remark now that while the black subgraph of both our automata (\CTA\ and its stratified version) can have any form, the red one is quite specific. 
To describe its components, we introduce a new class of TA, \emph{speedy} ones. Such TA exhibit ``obese-like'' behavior during a time interval.

\begin{definition}\label{def:speedy}
Given
%  a  proper subset of clocks 
$Z\subsetneq X$, we say that a strongly connected \RTA\ $\D$ with at least one transition is $Z$-\emph{speedy}  (or just \emph{speedy}) whenever
    clocks  not in  $Z$ are never reset;
    and all the guards have the form (for some constants $d_y$)
    $$\guard_\D =\bigwedge_{z\in Z}(0<z<1)\land \bigwedge_{y\not\in Z}(d_y<y<d_{y}+1).$$
The set of  all clocks reset by transitions in $\D$ is denoted $\reset_\D\subseteq Z$.
\end{definition}
% Given a $Z$-speedy automaton $\D$, let $\reset_\D\subseteq Z$ be the set of all clocks reset by transitions in $\D$. We consider  its  support $\supp \D$ and trivially timed version $\widetilde{\D}$
% %, as described in \cref{con:supp},
%  and state an analog of \cref{prop:trivial}.% with all clock constrains, i.e.~ guards and starting conditions,   removed (its transitions are all $\widetilde{\delta}$, such that $\delta=(p,a,\guard,\reset,q)$ is a transition of $\D$ and $\widetilde{\delta}=(p,a,\top,\reset,q)$).
% %
Let us characterize the size of languages of speedy automata.
\begin{restatable}{lemma}{lemspeedylight}\label{lem:speedy:light}
Let $\D$ be a speedy \RTA,  with   $\alpha = \gro\squeeze (\supp \D)$. Let  $L$ be the timed language of $\D$ with some nonempty sets of initial and final locations. Then for $t\in(0,1]$    the language $L_t$ has  entropy and capacity $\ent_{\varepsilon/2}(L_t) \eqapprox  \capa_{\varepsilon} (L_t) \eqapprox \frac {\alpha t}\varepsilon$.
\end{restatable}
The proof is based on a comparison of $\D$ with its trivially timed version $\widetilde{\D}$.
%and uses \cref{prop:trivial}. 
%
%The need for speedy automata is justified by 
The following lemma justifies speedy automata:
\begin{restatable}{lemma}{lemredispeedy}\label{lem:red:is:speedy}
In an \CTA, any red SCC is $Z$-speedy (with $Z$ the set of clocks being reset within it). 
In a stratified automaton, in every red SCC,  all locations are avatars $(q,Z)$ with a same $Z$, and such SCC is $Z$-speedy.
\end{restatable}
 %
% The latter speedy red SCCs will play a key role in the bandwidth computation, and we will give them a name.
\begin{definition}Red SCCs in a stratified automaton are called
   \emph{spots},  the set thereof is denoted $\Dset$. The \emph{growth rate} of a spot $\D$ is defined as $\alpha_\D = \gro\squeeze (\supp \D)$.  
\end{definition}

% In conclusion, we can associate, to each spot, the logarithm of an algebraic number, computed as the spectral radius of the 
% adjacency matrix of a deterministic finite automaton, which corresponds with the determinization of the squeezing of the support of the spot. 
% Each such value, which we will subsequently refer to as the \emph{growth rate of the spot},
% will be used in the following abstraction step, producing a weighted timed automaton. 
 
% Our bandwidth-preserving pre-treatment of the TA is over, we have it now in the stratified form, all spots are identified and their growth rates are computed. We can now proceed with the abstraction step. 
 
%\catalin{rephrase si tu veux -- non je veux pas !}

\subsection{Abstraction of a stratified automaton}\label{sec:abs}
% The algorithm for computing the bandwidth of a stratified obese TA $\aut$ is based on its transformation into an ``abstract'' WTG. The idea is to factorize behaviors of a stratified automaton into several speedy timed modes (corresponding to spots) and transient processes between them. A run through each of the spots is replaced by a sojourn in a state whose weight is the growth rate of this spot, followed by a unique ``abstract'' event.  We will describe the transformation in detail and prove that the \textbf{bandwidth} of the obese automaton equals the maximal \textbf{reward-to-time ratio} of long runs in the abstract timed graph, computable as established in \cite{alive}; see our \cref{sec:alive}.

We transform now the stratified automaton $\aut$ into an ``abstract'' WTG $\widehat{\aut}$ by processing (abstracting) every spot $\D$. All red transitions within $\D$ are removed and runs  within it are replaced by first waiting in a state of weight $\alpha_D$ and next taking a special abstract transition. All other states weigh $0$. 
More precisely: 
\begin{con}\label{con:abstract}
 The stratified automaton $\aut=(Q,X,\Pset{\Sigma},\Delta,S,I,F)$,  is transformed into the \emph{abstract} WTG
$\widehat{\aut}=(\widehat{Q},X,\widehat{\Delta},\widehat{S},w)$ with
\begin{itemize}
    \item $\widehat{Q}= Q\cup \check{Q}$ with $\check{Q}=\{\check{q}: q\in\cup\Dset\}$, i.e., we add a copy $\check{q}$ (``abstract location'') for each $q$ within a spot (to represent entering and evolving in this spot)
    % \item $\widehat\Sigma=\Pset{\Sigma}\cup \{\pqd : \D\in \Dset \land p,q \in \D \}$, that is, we add to the alphabet an abstract letter for each spot and pair of its states.
  %  $\widehat(\Sigma)=\Sigma\cup\Dset$, that is we add to the alphabet an abstract letter for each spot.
    \item $\widehat\Delta$ is constructed as follows: 
       \begin{itemize}
           \item all the transitions in $\Delta$  except red ones within the same spot are preserved (labels are removed); 
           \item for a red transition $p\trans{a,\guard,\reset}q$ with $p\not\in \D$ and $q\in \D$  for some spot $\D$, and for  a black transition $p\trans{a,\guard,\reset}q$ with  $q\in \D$, a copy  $p \trans{\guard,\reset}\check{q}$ is created; 
           %\item similarly for  each black transition $p\trans{a,\guard,\reset}q$ with  $q\in \D$   a copy  $p \trans{a,\guard,\reset}\check{q}$ is created;
           \item for each spot $\D$ and   locations $p,q\in \D$, a new abstract transition $\check{p} \trans{\guard_\D,\reset_\D}q$ is introduced, with $\guard_\D$ and $\reset_\D$ as in \cref{def:speedy}.
       \end{itemize}
    \item For $q \in Q$, the starting conditions are  $\widehat{S}(q)=\widehat{S}(\check{q})=S(q)$.
    \item For all locations $p$ in a spot $\D$  we put  $w(\check p)= \alpha_D$, which is the growth rate of the spot. All other weights in $\widehat{\aut}$ are $0$.
\end{itemize}
\end{con}
%Informally, all red transitions within a spot $\D$ are removed and replaced by a few abstract transitions from each ``entry'' location $\check p$ to each ``exit'' location $q$. 
An important property of  the abstract WTG $\widehat\aut$ is its time-divergence, and thanks to it, \cref{lem:alive} can be applied to compute the optimal reward-to time ratio. We will show that this ratio equals the bandwidth coefficient that we aim to compute.

%

% \begin{proof}
% The former statement follows from \cref{lem:nored:nonobese}, the latter from \cref{lem:black:non0}.
% \end{proof}
\subsection{Main result and an example of computation}\label{sec:example}
Let $\alpha\in\Log$ be the maximal reward-to-time ratio for $\widehat\aut$ (computable from $\aut$).
\begin{thm}
Either $\alpha=0$ and $\aut$ is not obese, or
$\band\capa_\varepsilon(\aut)\eqapprox\band\ent_{\varepsilon/2}(\aut)\eqapprox \alpha/\varepsilon.$
\end{thm} 
\begin{proof}[ideas]
To prove that $\band\ent_{\varepsilon/2}(\aut)\eqapprox \alpha/\varepsilon$ we consider a small $\varepsilon/2$-net  $\Net$ for all runs in the WTG $\hat A$ of duration $\leq T$ and build an $\varepsilon/2$-net from it for $L_T(\aut)$ by replacing, in each run, each abstract transition by an $\varepsilon/2$-net for the corresponding speedy language (with  growth rate at most $\alpha/\varepsilon$).

% To prove that $\band\ent_{\varepsilon/2}(\aut)\eqapprox \alpha/\varepsilon$ we consider a small $\varepsilon/2$-net  $\Net$ for all runs in the abstract timed graph $\hat A$ of duration $\leq T$. In each element $\rho=(\delta_1,d_1)\dots (\delta_n,d_n)$  of this net we replace each concrete transition  $\delta_i$ by its label in $\aut$ with the same delay $d_i$, and each abstract transition  $\delta_j$ by the $\varepsilon/2$-net for the corresponding speedy language of duration $d_j$  (its size is $\lesssim 2^{\alpha_j d_j/\varepsilon}$, where $\alpha_j$ is the bandwidth of the corresponding speedy spot, due to \cref{lem:speedy:light}). It is not difficult to show that this construction yields  a set $\Net(\rho)$ of cardinality $$\lesssim 2\strut^{\sum\alpha_j d_j/\varepsilon}= 2^{w(\rho)/\varepsilon}\lesssim 2^{\alpha T/\varepsilon}.$$ The union of $\Net(\rho)$ over all $\rho\in\Net$ provides a required  $\varepsilon/2$-net for $L_T(\aut)$.  

Towards the opposite inequality, we take in the graph $\hat A$ a run \linebreak$(\delta_1,d_1)\dots (\delta_n,d_n)$ of duration $T$ with almost optimal reward-to-time ratio. We replace each abstract transition  $\delta_j$ by the $\varepsilon$-separated set in the corresponding speedy language of duration $d_j$  (its size is $\gtrsim  2^{\alpha_j d_j/\varepsilon}$ due to \cref{lem:speedy:light}).\qed

% Towards the opposite inequality, we take in the graph $\hat A$ a run $(\delta_1,d_1)\dots (\delta_n,d_n)$ of duration $T$ with almost optimal reward-to-time ratio. We replace each concrete transition  $\delta_i$ by its label in $\aut$ with the same delay $d_i$, and each abstract transition  $\delta_j$ by the $\varepsilon$-separated set in the corresponding speedy language of duration $d_j$  (its size is $\gtrsim  2^{\alpha_j d_j/\varepsilon}$ due to \cref{lem:speedy:light}). This yields an $\varepsilon$-separated set  in  $L_T(\aut)$ of cardinality $\gtrsim 2\strut^{\sum\alpha_j d_j/\varepsilon}= 2^{w(\rho)/\varepsilon} \gtrsim 2^{\alpha T/\varepsilon}$.\qed
\end{proof}
%
%The algorithm for computing bandwidth, summarizing all the results in this section, is presented in \cref{fig:algos}. 
An \textbf{example of bandwidth computation} given in \cref{fig:avatars} is slightly simplified (with respect to the algorithm) for the sake of clarity. % The automaton considered (left) is not, technically speaking, a \CTA, but it is 0-free, and thanks to the clock $y$  (present in states $p$ and $q$), it satisfies the conclusion of \cref{lem:heart}. In the stratified automaton in the middle, we do not represent the sink state $(p,\emptyset)$. 
Pink rectangles in the stratified automaton correspond to spots (to be abstracted at the next step). Growth rate of the upper spot is $\log((7 + \sqrt{57})/2)\approx 2.863$ (as shown in \cref{sec:squeeze}), it is $2$ and $3$ for the  two others.  In the abstract WTG, all the red arrows within each spot are replaced by an abstract blue transition from the input to the output states. Weights of input states correspond to the growth rates of their respective spots.   The optimal cycle in the WTG has duration $7$ and spends one time unit in both $\check{p}$ and $\check{r}$, with the ratio $\approx (3+2.863)/7\approx 0.838$. This is better than the slightly faster cycle of duration $6$ spending one time unit in both $\check{q}$ and $\check{r}$, with the ratio $(2+3)/6\approx 0.833$.
Thus the bandwidth is $\approx0.838/\varepsilon$.

\section{Conclusions} \label{sec:con}

%\catalin{Pas plus de 2 paragraphes}

We have shown that the bandwidth of an obese TA can be computed as the logarithm of an
algebraic number, similarly to finite automata. The algorithm for computing it requires a number of
preprocessing steps: splitting into regions, dealing with $0$-transitions
(which required introducing a generalization of the pseudo-distance in \cite{distance}),
separating ``high-frequency'' components (called here \emph{spots}) from ``normal-frequency'' components,
computing the bandwidth of spots, 
transforming the timed automaton into a WTG where weights are the bandwidths of the various spots,
and finally applying results from \cite{alive} for computing the maximal reward-per-time ratio
of the resulting WTG.% using its finite corner-point abstraction. 

As for complexity, we remark that by reduction from reachability, similar to that of \cite[Thm.~6]{3classes}, deciding whether $\alpha=1$ or $2$ (even knowing that one of those holds), is \textsc{PSPACE}-hard.  We leave a more precise complexity analysis as a future work.  
%
%For future work, 
We also plan to address the problem of computing the bandwidth of  the last remaining class of timed automata, normal ones, by characterizing the coefficient $\alpha$
such that $\bandc_\varepsilon(L(\aut))\eqapprox \alpha\cdot \log(1/\varepsilon)$ 
 as the best reward-to-time ratio of some finite abstraction of the timed automaton. We  intend to apply these results to the approximate coding and compression of timed data.

\bibliographystyle{splncs04}
\bibliography{entro}

 \appendix
 \section{Preliminary details}

\subsection{More on timed automata}

\begin{definition}[\cite{alive}, variant]
%[Bouyer et al., variant]
Given a bounded TA (or graph), a \emph{corner-point} is either a tuple $(q,R,\x)$ with $q\in Q$,  $R\subseteq S(q)$ a region, and $\x$ a vertex of $R$, or a tuple $(\delta, R, \x)$ with $\delta\in\Delta$, $R\subseteq \guard_\delta$ a region, and $\x$ a vertex of $R$.
\end{definition}

\begin{con}[also \cite{alive}]\label{con:corner}
The \emph{corner-point graph} of a bounded TA (or graph)  is the finite graph whose vertices are all the corner-points,  having two kinds of edges:
\begin{description}
    \item [delay edges:] from any corner-point $(q,R,\x)$ to any $(\delta,R',\x')$ such that $\delta$ is of the form $(q,?,?,?,?)$ and $\x'=\x+t$ for some  $t\in\nat$, we call $t$ the duration of the edge;
    \item [jump edges:] from any corner-point $(\delta,R,\x)$ to any  $(q',R',\x')$ such that $\delta$ is of the form $(?,q',?,?,\reset)$ and  $\x'=\x[\reset]$, its duration is $0$.
\end{description}
\end{con}

% \subsection{Notations on asymptotic bounds}\label{sec:not}
% For $\xi \in \real$, we denote $\lfloor \xi \rfloor$ its integral part
% and
% $\{\xi\}$  its fractional part. All logarithms in this article are base 2.
% We denote  by $\Log$ the set of logarithms  of positive algebraic numbers.
% %
% Asymptotic results will be stated using the following notation:
% \begin{definition}
% Given  two positive-valued  functions $f(\varepsilon,x)$ and $g(\varepsilon, x)$ (with any kind of arguments $x$), we write $f\lesssim g$ whenever 
% $$
% \forall \varkappa>0\; \exists c,\varepsilon_0>0,\forall\varepsilon<\varepsilon_0,\forall x: \, f(\varepsilon,x)< c  g^{1+\varkappa}(\varepsilon,x).
% $$
% %
% % for all $\varkappa>0$ exist $c$ and $\varepsilon_0>0$ such that for all $\varepsilon<\varepsilon_0$ and all $x$ it holds that $f(\varepsilon,x)< c  g^{1+\varkappa}(\varepsilon,x)$.  
% %
% For real-valued  $f(\varepsilon,x)$ and $g(\varepsilon, x)$, we write $f\lessapprox g$ whenever 
% $$
% \forall \varkappa>0\; \exists c,\varepsilon_0>0,\forall\varepsilon<\varepsilon_0,\forall x: \, f(\varepsilon,x)< c+ (1+\varkappa)g(\varepsilon,x).
% $$
% We write $f\eqsim g$ if $f\lesssim g$ and $g\lesssim f$, and $f\eqapprox g$ if $f\lessapprox g$ and $g\lessapprox f$.
% \end{definition}
% In more usual terms, $f\lessapprox g$ iff $f\leq g +o(g)+O(1)$ as $\varepsilon\to 0$ and uniformly in other parameters.

\subsection{More on notation} 

\begin{proposition}\label{prop:log}The set $\Log$ has the following properties:
\begin{itemize}
    \item for $x,y \in \Log$ and $c\in \rat$, it holds that $x+y, cx\in\Log$;
    \item numbers in $\Log$ admit a finite representation, such that equality, comparison, addition, and multiplication with a  rational are all computable;
    \item given $x\in\Log$ and $n\in \nat$, a rational approximation of $x$ with precision $2^{-n}$ is computable\footnote{i.e.~$\Log$ is an effective subset of computable real numbers \cite{analysis}}.
\end{itemize}
\end{proposition}

\begin{proposition}
Relations $\lesssim,\lessapprox$ are transitive and enjoy the following properties:
    \begin{align*}
        f\lesssim g &\Leftrightarrow  \log f \lessapprox \log g;\\
        f\lesssim f' \land  g\lesssim g' &\Rightarrow  fg\lesssim f'g';\\
        f\lessapprox f' \land  g\lessapprox g' &\Rightarrow  f+g\lessapprox f'+g';\\
        f_i\lessapprox g \text{ uniformly } &\Rightarrow \sum_{i=1}^h f_i \lessapprox g h + Ch \text{ for some  } C.  
    \end{align*}
\end{proposition}

\begin{definition}
   For various kinds of automata, we will denote  by $\pql(\aut)$ the language corresponding to the initial state (or location) $p$ and final $q$ and by $L_{\full}(\aut)$ the language where all states are initial and final. 
\end{definition}

\subsection{More on reward-to-time ratio }\label{sec:alive:bis}
Results in \cite{ratio} can be summarized as follows,

\begin{lemma}[see \cite{ratio}]
  Given a finite directed graph  with non-negative rewards and  times associated to edges, and sum of times $\geq 1$ for every cycle, let $\alpha>0$ be the maximal reward-per-time ratio over all simple cycles. Then  for all paths $\pi$, 
$
w(\pi)\lessapprox\alpha t(\pi).
$
\end{lemma}
A remark on algorithmic aspects is in order. We will consider graphs with integer times and weights in $\Log$. %Due to \cref{prop:log} 
Hence, the cycle ratios are also in $\Log$. It follows that the brute force algorithm enumerating all the simple cycles and comparing their ratios would find the exact value of the maximal ratio as a member of $\Log$.

 Alternatively, the computation of a numeric solution with $k$ decimals would be polynomial in $k$ and the size of the graph, using one of the classical algorithms.

% %\eugene{check}

Let us give more details on results and algorithms in  \cite{alive}. The authors of \cite{alive} associate a WTG to its weighted corner-point graph, proceeding as in \cref{con:corner}, then assigning  reward and time  0 to all jump edges. Finally, for each delay edge $e$ from location $p$ with  duration $t$, they put $t(e)=t$ and $w(e)=w(p)t$. 

\cref{lem:alive} follows from the stronger result
\begin{lemma}[\cite{alive}, variant]\label{lem:alive:bis}
Given a bounded  time-divergent WTG, let $\alpha>0$ be the maximal reward-per-time ratio over all simple cycles in the corner-point graph. Then there exists a constant $C$ such that 
\begin{itemize}
    \item for all runs $\rho$,
$
w(\rho)\leq\alpha \length(\rho)+C;
$
\item   for any  $T$ there exists a run  $\rho$ with $\length(\rho)\leq T$ and $w(\rho)>\alpha T -C$.
\end{itemize}
\end{lemma}
The former follows from the proof of \cite[Prop.~4]{alive},  the latter from  
\cite[Prop.~5]{alive}.

The algorithm  in \cite{alive} for computing $\alpha$ consists in constructing the corner-point graph, and computing the reward-to-time ratio for it

\subsection{More on squeezed languages}

For the following, we recall Fekete's Lemma:
\begin{lemma}[\cite{fekete}] If a function $g:\nat\to\real$ satisfies the superadditivity condition $g(m+n)\geq g(m)+g(n)$ then the limit $\lim_{n\to \infty}g(n)/n$ (finite or infinite) exists.
\end{lemma}

We are interested in a specific case of strongly connected automata. %We vary initial and final sets $I$ and $F$ and define $\pql^\aut$ as the  language accepted by $\aut$ with $I=\{p\}$ and $F=\{q\}$, and also $L_{\full}^\aut$ as the accepted language with $I=F=Q$.

\begin{proposition}\label{prop:SCC:squeez} Let $\aut=(Q,\pset\Sigma,\Delta,I,F)$ a strongly connected finite automaton with non-empty $\Delta,I,F$ and $\alpha=\gro{\squeeze (L(\aut))}$. Then the following holds:
\begin{enumerate}
\item $\alpha>0$.
   \item \label{item:pq} $\#\squeeze_n \pql^\aut \eqsim \#\squeeze_n L_{\full}^\aut$ for all $p,q\in Q$ and $n\in \nat^+$.
     %\item The growth rate does not depend on the choice of non-empty $I,F$.
     \item \label{item:lim}The growth rate is in fact a limit:  $\alpha=\lim_{n\to\infty} \frac{\log\#\squeeze_n L_\full}{n}$.
     \item  $ \#\squeeze_n L_\full^\aut\lesssim 2^{\alpha n}$ for all $n\in \nat$.
     \item $2^{\alpha n}\lesssim  \#\squeeze_n \pql^\aut$ for all $p,q\in Q$ and $n\in \nat^+$.
 \end{enumerate}

% \begin{enumerate}
%     \item For some $c_1>0$, it holds  that $c_1\leq \#\squeeze_n (\pql^\aut)\big/ \#\squeeze_n (L_{\full}^\aut) \leq 1$ for all $n\in \nat^+$.
%     \item The growth rate does not depend on the choice of non-empty $I,F$
%     \item The growth rate is in fact a limit:  $\rho=\lim_{n\to\infty} \frac{\log\#\squeeze_n L_\full}{n}$.
%     \item For any $\varkappa>0$ exists $c_2>0$ such that $ \#\squeeze_n (L_\full^\aut)\leq c_2 2^{(\rho+\varkappa) n}$ for any $n\in \nat$.
%     \item For any $\varkappa>0$ exists $c_3>0$ such that  $ c_3 2^{(\rho-\varkappa) n}\leq \#\squeeze_n (\pql^\aut)$ for all $p,q\in Q$ and $n\in \nat^+$.
% \end{enumerate}
\end{proposition}
%\eugene{could we do better/simpler using Perron-Frobenius?}
\begin{proof} Let $U\subseteq \Sigma$ be the union of all transition labels in $\aut$.
We remark that in the automaton  $\squeeze \aut$, there exists a transition from each state $p$ to each state $q$  labeled by $U$. Indeed, there is a run in $\aut$ from $p$ to $q$ on some word $v$, including all the labels due to strong connectedness. Hence, there is a transition $p\trans{U}q $ in $\squeeze\aut$.
\begin{enumerate}
\item  %\eugene{to prove}
Since $\aut$ is strongly connected (and trim), its language contains a sublanguage of the form $uv^*$ with $v$ not empty. This means that $\squeeze L(\aut)\supseteq u(\emptyset+\Let(v))^*$. But the growth rate of the right-hand side term is $1$.
    \item  Trivially $\squeeze_n \pql^\aut\subseteq\squeeze_n L_{\full}^\aut$ which implies the $\lesssim$ inequality.
    
    We remark now that every word $W \in \squeeze L_{\full}^\aut$  yields $UWU \in \pql^\aut$. Indeed,  suppose that  $W$ is accepted by a run from $r$ to $s$ in $\squeeze\aut$. Then exists a run  $p\trans{U} r\Trans{W}s\trans{U}q$. Thus we have an injection $W\mapsto UWU$, from $\squeeze_{n-2}L_{\full}^\aut$ to $\squeeze_n \pql^\aut$, and $\#\squeeze_{n-2}L_{\full}^\aut\leq \#\squeeze_n \pql^\aut$. It is easy to see that $\#\squeeze_{n}L_{\full}^\aut\leq 2^{2\#\Sigma}\#\squeeze_{n-2}L_{\full}^\aut$, and we can conclude that $\#\squeeze_{n}L_{\full}^\aut\lesssim \#\squeeze_n (\pql^\aut)$.

    %\item Immediate from the previous statement.
    \item Let $f(n)= \# \squeeze_n L_{\full}^\aut$. We will prove that $f$ is almost supermultiplicative.   Indeed, given two words $V$ and $W$  of $m$ and $n$ letters respectively, accepted by runs from $p$ to $q$ and from $r$ to $s$, there is a run $p\Trans{V} r\trans{U}s\Trans{W}q$. Then the mapping $V,W\mapsto VUW$ is an injection from $\squeeze_m L_{\full}^\aut \times  \squeeze_n  L_{\full}^\aut$ to  $\squeeze_{m+n+1} L_{\full}^\aut$.  This injection implies that $f(m+1+n)\geq f(n)f(m)$  which, in turn, implies the existence of $\lim_{n\to\infty}\log f(n)/n$  by  applying Fekete's Lemma to $g(n)=\log f(n-1)$.
    
    \item Immediate from the previous statement.
    \item Immediate from  the statements \ref{item:pq}, \ref{item:lim}. \qedhere
\end{enumerate}
\end{proof}

\subsection{On trivially timed languages}
\proptrivial*

We will provide  and prove a more detailed version of this proposition:
\begin{proposition} \label{prop:trivial:bis}In the hypotheses of \cref{prop:trivial}
\begin{itemize}
    \item  there exists an $\varepsilon/2$-net $\Net_{T,\varepsilon}$ of size $\lesssim 2^{\alpha T/\varepsilon}$ for  $L_{T,\full}(\aut)$;
   \item if $T>2\varepsilon$, then for each two states $p,q$ there exists an $\varepsilon$-separated set $\Sep_{T,\varepsilon}^{pq}$ of size $\gtrsim 2^{\alpha T/\varepsilon}$ within  $\pql_T(\aut)$. Moreover, elements of $\Sep$ contain no events on $(0,\varepsilon]$ nor $[T-\varepsilon,T)$.
\end{itemize}
\end{proposition}
\begin{proof}
We sketch first the construction of $\Sep$. Let $K=\lceil T/\varepsilon\rceil-1$ and $0=t_0<t_1<\cdots<T_K=T$ be equidistant points in $[0,T]$ (with step slightly $>\varepsilon$).

The squeezed language $ \squeeze_{K-1}\pql(\supp\aut)$ contains $\eqsim 2^{\alpha (K-1)}\eqsim  2^{\alpha T/\varepsilon}$ distinct words. By definition of squeezing, each such word has the form
$W=\lett{w_1}\lett{w_2}\dots \lett{w_{K-1}} $ for some $w_1 w_2\dots w_{K-1}\in \pql^\aut$.  Let $\beta(W)$ be the timed word containing each $w_i$ at time $t_i$ (for $i=1..K-1$). It is easy to see that $\beta(W)\in \pql_T^\aut$. Given $W\neq W' \in \squeeze\pql(\supp\aut)$, necessarily $\lett{w_i}\neq \lett{w'_i}$, for some $i$, that is some letter $a$ present at $w_i$ but not at $w'_i$ or vice versa. Hence, $a$ takes place at $t_i$ in one of the two timed words $\beta(W), \beta(W')$ but not in the other, and there are no matching characters in the radius $\leq \varepsilon$. We conclude that $\beta(\squeeze_{K-1}\pql(\supp\aut))$ is an $\varepsilon$-separated set within  $\pql_T^\aut$ of size $\eqsim  2^{\alpha T/\varepsilon}$ as required.

To build $\Net$, we proceed similarly. Choosing $K=\lceil T/\varepsilon\rceil$ we get a grid $0=t_0<t_1<\cdots<T_K=T$ with step slightly $\leq\varepsilon$. We take any  word $w=\lett{w_0}\lett{w_1}\dots \lett{w_{K}}$ in the squeezed language $ \squeeze_{K+1}L(\supp\aut_\full)$, position each $w_i$ at time $t_i$ (for  $i=0..K$),  and call the result $\gamma(w)$. The image  
$\gamma(\squeeze_{K+1}(\supp\aut_\full))$ is the required $\varepsilon/2$-net of size $\eqsim  2^{\alpha T/\varepsilon}$. Indeed, for any timed word  $v\in L_{T,\full}(\aut)$,  we can obtain an $\varepsilon/2$-close $v'$ by moving each letter to the closest grid point $t_i$. Let $V_i$ be the set of letters in $v'$ at $t_i$, and let $V=V_0V_1\dots V_K$, then $v''=\gamma(V)$ contains exactly the same letters as $v'$ at each $t_i$, hence $d(v'',v')=0$ and, by triangular inequality $d(v'',v)\leq\varepsilon$, as required. \qed
\end{proof}

\section{Proofs on transformation of automata}
\subsection{Closure of \RTA}

We  introduce the notation $\overline{\aut}$ for the closed version of any \RTA\ $\aut$: the same automaton where all strict inequalities are changed to large inequalities. Thanks to region-splitting, the closure retains the same set of symbolic behaviors as the original \RTA: for a given path, any run along this path in the closure can be matched to a run along the same path in the original automaton (and it can be chosen with arbitrarily close to identical delays). This allows us to safely reason on runs that visit vertices and faces of regions.% \eugene{correct but cryptic, clarify please} \aldric{mieux comme ça ?}
%\eugene{faire un vrai lemme d'Aldric}

More formally, the following relations between the path of a \RTA\  and its closure holds:

\begin{lemma}[\cite{3classes}]\label{lem:closed-semantics}
In an \RTA, for any  path $\pi$ and clock vectors $x,y$:
\begin{itemize}
 \item $L_{\pi^*,T}(x,y)\subseteq L_{\bar{\pi}^*,T}(x,y)$;
 \item $\ent_\varepsilon(L_{\pi^*,T}(x,y)) = \ent_\varepsilon(L_{\bar{\pi}^*,T}(x,y))$;
 \item $\capa_\varepsilon(L_{\pi^*,T}(x,y)) \leq \capa_\varepsilon(L_{\bar{\pi}^*,T}(x,y))$;
 \item for any $0<\varepsilon'<\varepsilon$, $\capa_{\varepsilon'}(L_{\pi^*,T}(x,y)) > \capa_{\varepsilon}(L_{\bar{\pi}^*,T}(x,y))$.
\end{itemize}
\end{lemma}

This translates directly in terms of bandwidth:

\begin{corollary}\label{cor:closed}
Let $\aut$ be an \RTA. For all $0<\varepsilon'<\varepsilon$, we have $$\band\capa_\varepsilon(L(\aut))\leq \band\capa_\varepsilon(L(\overline{\aut}))\leq\band\capa_{\varepsilon'}(L(\aut)).$$
\end{corollary}

%To facilitate reasoning on corner-points, we make a preliminary observation: 
This construction also makes it possible to obtain stronger statements in several cases. For instance, the following observation will facilitate reasoning on corner-points:

\begin{lemma}\label{lem:closed-duration}
A path $\pi$ in a $\RTA$ can be realized in time $<1$ if and only if its closure $\bar\pi$ can be realized in time $0$.
\end{lemma}

\begin{proof}
It is a convex optimization problem: we optimize total duration by varying individual delays, respecting closed timed path constraints. The optimal solution is realized on a vertex of the polyhedron of delays, but this vertex corresponds to a path along region vertices, hence with integer delays. So the optimal total duration for closed constraints is an integer, necessarily smaller than any total duration obtained for open constraints; hence, if $\pi$ can be realized in duration $<1$, this integer is $0$. Conversely, if $0$ is the optimal duration, then you can just take any run, having delays in the inside of the delay polyhedron and close enough to those of a path realizing $0$, such that its duration is $<1$. \qed
\end{proof}

\subsection{Properties of heartbeat, zero-elimination and region-splitting}

\begin{lemma}\label{lem:heart:band}
Adding the heartbeat does not change the bandwidth.
\end{lemma}
\begin{proof}[sketch]
Adding or erasing $b$ every time unit transforms an $\varepsilon$-net into $\varepsilon$-net and an $\varepsilon$-separated set into an $\varepsilon$-separated set of the same size.
\end{proof}

To reason about 0-elimination, we need a couple of definitions

\begin{definition}[0-elimination, words]\label{def:nu} A timed word $w = (a_1, t_1)\dots (a_n, t_n)$ is \emph{0-free} whenever $0<t_1<t_2<\cdots <t_n$.
Given a timed word $w$ over an alphabet $\Sigma$, its \emph{$0$-elimination} is the 0-free timed word $W=\NoZ(w)$ over the alphabet $\Pset{\Sigma}$  obtained as follows:
\begin{itemize}[nosep]
    \item let $t_1<t_2<\cdots<t_n$ be all the distinct non-zero event times in $w$;
    \item let $A_i\in \Pset{\Sigma}$ be the set of all the events in $w$ occurring at time $t_i$;
    \item finally let $W=(A_1,t_1),(A_2,t_2),\dots, (A_n,t_n)$.
\end{itemize}
\end{definition}
For example $\NoZ((b,0)(a,5)(b,5)(a,5)(c,7))=(\{a,b\},5)(\{c\},7)$.

\begin{lemma}\label{lem:0:band}
Operation $\NoZ$ extended to timed languages preserves both $\bandh$ and $\bandc$.
\end{lemma}

\begin{proof}[sketch]%\todo{make it a proof}
We recall that $\ell(u)$ denotes the set of all letters in a timed word $u$. The other way around for $A\in\pset{\Sigma}$ let $\sigma(A)$ be the timed word in $\Sigma$ consisting of all letters in $A$ (in alphabetic order) at time $0$.

\textbf{Let $N$ be an $\varepsilon$-net for $L$.} % We decompose each $w\in N$  as $uv$, where $u$ are all letters at time $0$. Let now $w'$ be $\NoZ(v)$. 
%with $(\ell(u),\varepsilon/2)$ inserted. 
For each $w\in N$ and each $A\in\pset{\Sigma}$, the new set $N'$ contains  $\NoZ(w)$ with $(A,\varepsilon/2)$ inserted. It is an $\varepsilon$-net for $\NoZ(L)$ and $\#N' \leq 2^{\#\Sigma}\# N$. 

\textbf{Let  $N$ be an $\varepsilon$-net for $\NoZ(L)$.} For each $w\in N$ and each $A\in\pset{\Sigma}$ we replace each letter $B$ in $w$ by $\sigma(B)$, and insert $\sigma(A)$ at time $0$. the set $N'$ of all timed words obtained in this way is an $\varepsilon$-net for $L$, and  $\#N'\leq 2^{\#\Sigma}\# N$.

\textbf{Let $S\subseteq L$ be an $\varepsilon$-separated set.} We decompose each $w\in N$  as $uv$, where $u$ are all letters at time $0$, and choose  the most  frequent value  of $\ell(u)$, we denote it by $A\in\pset{\Sigma}$. Let $N'$ be the set of all $w\in N$ starting with $u$ such that  $\ell(u)=A$. Then $\NoZ(N')\subseteq \NoZ(L)$ is $\varepsilon$-separated  and 
$\#\NoZ(N')\geq 2^{-\#\Sigma}\# N$. 

\textbf{Let finally $S\subseteq \NoZ(L)$ be an $\varepsilon$-separated set.} For each $w\in N$  we choose some $w'\in L$ such that $w=\NoZ(w')$, and decompose $w'$ as $xyv$, where $x$ are all letters at time $0$, and $y$ all letters at time $(0,\varepsilon]$, and choose  the most  frequent value  of $(\ell(x),\ell(y))$, we denote it by $A,B\in\pset{\Sigma}^2$. Let $N'$ be the set of all $w'\in L$, corresponding to $w\in N$, and such that  $\ell(x)=A$ and $\ell(y)=B$. Then $N'\subseteq L$ is $\varepsilon$-separated and $\#N'\geq 2^{-2\#\Sigma}\# N$.

In all cases, sizes of $\varepsilon$-nets and separated sets for $L$ and $\NoZ(L)$ differ only by a multiplicative constant, and the assertion to prove follows. \qed
\end{proof}

\begin{figure}[t]\scriptsize
\begin{tikzpicture}
\node[state](p)[initial]  at (0,0)  {$p$};
\node[state](q) at (2.5,0) {$q$};
\node[state](r) at (4.5,1.3) {$r$};
\node[state](s) at (6,-0.5) {$s$};
\draw [post] (p) edge [above] node {$0<u<x<1$} node[below]{$a,\{u\}$} (q);
\draw [post] (q) edge [above,sloped] node {$0=u<x<1$} node[below,sloped]{$c,\{u\}$} (r);
\draw [post]  (r) edge [above,sloped] node {$0=u<x<1$}node[below]{$b,\{xu\}$} (s);
\end{tikzpicture}
\begin{tikzpicture}
\node[state](p)[initial]  at (0.5,0)  {$p$};
\node[state](q) at (4,0) {$q$};
\node[state](r) at (4,1.2) {$r$};
\node[state](s) at (4,-1.2) {$s$};
\draw [post] (p) edge [above] node {$0<u<x<1$}node[below]{$\{a\},\{u\}$ }(q);
\draw [post] (p) edge [bend left=7mm,above,sloped] node {$0<u<x<1$} node[below,sloped]{$\{ac\},\{u\}$}(r);
\draw [post] (p) edge [bend right=6mm,above,sloped] node {$0<u<x<1$}node[below,sloped]{$\{abc\},\{xu\}$} (s);
\end{tikzpicture}
\caption{0-elimination. Left: a fragment of \RTA\ with $S(p)=S(q)=S(r)=(0=u<x<1)$ and $S(s)=(x=u=0)$. Right: result of 0-elimination.
\label{fig:elim}}
\end{figure}
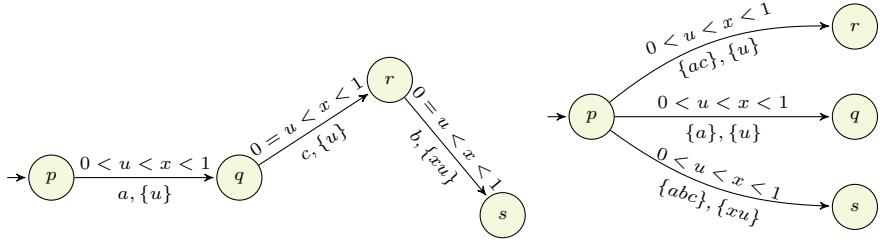

We illustrate \cref{con:0} by \cref{fig:elim}. 
Finally we remark that the automaton $\NoZ(\aut)$ provided by this construction recognizes $\NoZ(L(\aut))$, and due to \cref{lem:0:band}, has the same bandwidth as $\aut$. 

\subsection{About region-splitting}
Let us sketch the algorithm for bringing any timed  automaton with an upper bound on some clock at every transition into a \RTA\ form, see also \cite{AD,entroJourn}.
\begin{itemize}
    \item Eliminate all the ``diagonal'' constraints of the form $x \sim y+b$, as described in the proof of \cite[Lem.~4.2]{silent}.
    \item Bound all the clocks as in \cite[Lem.~31, ArXiV version]{3classes}.
    \item Proceed with splitting as in the proof of \cite[Prop.~8, ArXiV version]{3classes}.
    \item Observe that, due to upper bounds on clocks, all guards can be bounded, and split each guard into regions (this reintroduces diagonal constraints).
    \item Trim the obtained automaton.
\end{itemize}

\subsection{Properties of red transitions}
%
% The following description of red SCCs will play a key role in the proofs.
%
%\aldric{remarque~: ce n'est pas réellement un résultat sur le CPA mais plutôt sur les chemins de sommet dans la closure. Parler de CPA ici complique le message (force à parler de "counterparts" de "by contruction", etc.).}
Fast cycles can be characterized as follows:
\begin{lemma}\label{lem:black:non0}
A cycle $\pi$ is fast iff the corner-point graph has a cycle of duration $0$ along $\pi$.
%it has a run cycling in time $0$ through some vertex of its starting region.
\end{lemma}
\begin{proof}
In this proof, we use an alternative characterization of fast paths, which is a consequence of \cref{lem:closed-duration}: a path $\pi$ is fast if and only if $\bar\pi^2$ can be realized in time exactly $0$.

 Let $\pi$ be a cyclic path of an \RTA. By \cref{con:corner}, in its corner-point graph, any $0$-time path along $\bar\pi$, starting  from a corner-point $c$, goes to a corner-point $c[\reset_\pi]$, having the same coordinates as $c$ except on $\reset_\pi$ (the set of clocks reset in the transitions of the path $\pi$), where all coordinates are equal to $0$.
 
 Assuming $\bar\pi$ can be traversed twice in duration $0$ (i.e. $\pi$ is fast), then it can be done through a run $\rho$ along $\bar\pi^2$, entering and exiting each region through a corner-point (because the minimal duration can be realized on a vertex of the polytope of initial clocks and delays). It can be decomposed in this way: $\rho=\rho_1\rho_2$ where both $\rho_i$ are runs along $\bar\pi$. Let us call $c$ the starting corner-point of $\rho_1$. Then $\rho_2$ starts at $c[\reset_\pi]$, and thus $\rho_2$ ends at $c[\reset_\pi][\reset_\pi]=c[\reset_\pi]$. Hence $\rho_2$ is a cyclic run (of duration $0$ that enters and exits each region through a corner-point), so there exists a cycle of duration $0$, along $\pi$, in the CPA.
 
 Conversely, if a corner-point cycle of duration $0$, exists, it means, by construction, that a corresponding cyclic run $\rho$ of duration $0$ through region vertices existed in the closure of the original automaton. Let $\pi$ be its path, then $\bar\pi^2$ can be realized through the run $\rho^2$ of duration $0$, hence $\pi$ is fast. \qed
\end{proof}

%\begin{proof}[(Proof of old lemma)]
%One direction is straightforward: if $\pi$ is the run from the statement and $\rho$, the 0-time cyclic run through $\pi$, then $\rho^2$ is 0-time and goes through $\pi^2$.
%
%For the other direction, assume that no vertex admits a $0$-time cyclic run along $\pi$. Any $0$-time arrow of $\gamma(\pi)$ \aldric{(introducing timed orbits, sorry, do we copy/paste from 3-classes or introduce non-abstracted timed orbits? Nothing really justifies abstracting durations away here)}\eugene{probablement on devra rappeler le cornerpoint graph pour les résultats de Patricia. Est-ce suffisant ici?} goes from a vertex to the vertex having the same coordinates except on $\reset_\pi$, where the coordinate is 0. This implies that there cannot be two consecutive 0-time arrows (the second arrow would be a 0-time self-loop, which is excluded). Hence, there is no way to make a 0-time cycle of length $>2$ in the orbit, so $\gamma(\pi^2)$ contains no self-loop of time $0$.
%\end{proof}

From \cref{def:red} and the fact that heartbeat transitions are black and happen every time unit, we deduce:
\begin{lemma}\label{lem:heart}
Every run over red transitions stays in one red SCC and has a duration $<1$.
\end{lemma}

\lemredispeedy*
\begin{proof}
Let $\D$ be such an SCC. Let $Y$ be the set of clocks that are not reset along $\D$. First, remark that there is at least one such clock, the heartbeat clock $h$ (for resetting this clock, at least one black transition has to be taken). Because the automaton is in region-split form, for every location $q$, the region $S(q)$ satisfies $d_y(q)<y<d_y(q)+1$ or $d_y(q)=y$ for any $y\in Y$. Since the clock is not reset, it is not possible to have $d_y(q)\neq d_y(q')$ since it is possible to reach $q$ from $q'$ and $q'$ from $q$, which means $d_y(q)$ is the same for all locations in $\D$. Now, imagine the condition is of the form $d_y=y$ (it has to be as such for all locations). This means that any run through $\D$ has to be done in 0 time, but since the automaton is 0-free, this is not possible. This means $d_y<y<d_y+1$ for every starting region, which implies the condition is also satisfied by the guard of every red transition in $\D$.
Consider now the set of clocks $Z$ reset somewhere along $\D$. Necessarily, their fractional part is smaller than that of the heartbeat clock $h$, which is not reset. The clock $h$ being bounded by $1$, all clocks in $Z$ have the same bound as well. This means that the condition $z<1$ is satisfied for every transition in $\D$. The condition $0<z$ comes from 0-freedom. \qed
%
%Let $\D$ be such an SCC. Let $Y$ be the set of clocks that are not reset along $\D$. First, remark that there is at least one such clock, the heartbeat clock $h$ (for resetting this clock, at least one black transition has to be taken). The vertex $lowest(\D)$ gives $d_y$ for every such clock. Remark that if any such clock were to go beyond $d_y+1$ before a red transition in $\D$, then the arriving region would contain both $lowest(\D)$ and a vertex with $>d_y+1$ for clock $y$, which is not possible. This means the guard $d_y < z < d_y+1$ can be added to every red transition in $\D$ without changing the language.
%Consider now the set of clocks $Z$ reset somewhere along $\D$. Necessarily, their fractional part is smaller than that of the heartbeat clock $h$, which is not reset. $h$ being bounded by $1$, all clocks in $Z$ are as well. This means that the guard $z<1$ can be added for every red transition in $\D$.
\end{proof}

\subsection{On the stratified automaton}
\begin{lemma}
The timed languages of an \CTA\ and its stratified version coincide. 
\end{lemma}
\begin{proof}[sketch]
Let $\aut$ be an \CTA\ and  $\aut'$ its stratified version.
For $w\in L(\aut')$, if we take its accepting run and remove the avatar information (i.e.~map $((q,Z),\x)\mapsto (q,\x)$ for all the states), then we obtain an accepting run of $\aut$ on $w$, hence $L(\aut')\subseteq L(\aut)$.

The other way around, given $w\in L(\aut)$, consider its accepting run
$\rho=(q_0,x_0) \trans{(\delta_1,d_1)} (q_1,x_1)\cdots \trans{(\delta_n, d_n)}(q_n,x_n)$, and transform it into a run $\rho'$ of $\aut'$ as follows. Whenever $\delta_i$ is black, map $q_{i-1}$ to $(q_{i-1},\emptyset)$. If $\delta_i$ is red, let $Z$ be the set of clocks that are reset along the maximal contiguous red segment  of $\rho$ starting from $\delta_i$, and map $q_{i-1}$ to $(q_{i-1},Z)$. Finally, transform the last location into $(q_{n},\emptyset)$. It is easy to see that $\rho'$ is an accepting run of $\aut'$ on $w$. Hence $L(\aut)\subseteq L(\aut')$. \qed
\end{proof}

\subsubsection{A simulation relation on the stratified automaton}

We will use a partial order on clock configurations 
\begin{definition}[Partial order $\lessdot$]\label{def:lesss}
For any set $Z$ of clocks, we define a partial order on clock valuations as follows:
$$
\x\lessdot_Z x' \Leftrightarrow 
\left\{ 
\begin{array}{ll}
    c \in Z   &\Rightarrow x_c\leq x'_c;\\
   c \not\in Z  &\Rightarrow x_c = x'_c.
\end{array} 
\right.
$$
\end{definition}

We also define a partial order on states of $\aut'$:
\begin{definition}
For states of the stratified automaton, we say  that $s\lessdot s'$ whenever  $s=((q,Z), \x ), s'=((q,Z), \x')$ and $\x\lessdot_Z \x'$.
\end{definition}

We recall that a partial order $\mathcal{R}$ on states of a TA $\aut$ is a \emph{timed simulation} iff for all states $r, s, r'$ such that $r\trans{(\delta,t)} s$ and $r'\mathcal{R} r$, there exists a state $s'$ with $s'\mathcal{R} s$ such that  $r'\trans{(\delta,t)} s'$.

\begin{lemma}\label{lem:sim}Relation $\lessdot$ is a timed simulation on $\aut'$.
\end{lemma}
\begin{proof}
To prove the simulation property, let $r\trans{(\delta,t)} s$ be a transition and let $r'\lessdot r$. We need to find $s'\lessdot s$ such that $r'\trans{(\delta,t)} s'$. To that aim, we will consider two cases
\begin{itemize}
    \item If $r=(p,\emptyset)$, then $r'=r$, we take $s'=s$ and there is nothing to prove.
    \item It remains the non-trivial case of $r=(p,Z)$ with $Z\neq\emptyset$ and red $\delta$. Let $\D$ be the red SCC of $\delta$ in $\aut$, and all the transitions therein share the guard  $\guard_\D$ (due to \cref{lem:red:is:speedy}). By construction of $\aut'$
    $$
         r'= ((p,Z),\x'); \x'\lessdot_Z \x; s=((q,Z\setminus U),\y); 
          \x+t\models \guard_\D;  \y= \reset_\delta(\x+t); U\subseteq \reset_\delta\subseteq Z.
    $$
 Take now $\y'= \reset_\delta(\x'+t) $ and $s'=((q',Z\setminus U),\y')$. It is easy to check that  $\x'+t\models \guard_\D$; indeed, all the clocks $\not\in Z$ take the same values in $\x+t$ and $\x'+t$ and thus satisfy the same constraints. As for the clocks $i\in Z$  constrained to be $<1$, we have  $0<x'_i+t\leq x_i+t <1$, and the guard is satisfied. %\bernardo{Faut-il dire que x'+t>0 aussi ?} 
 
 Thus  $r'\trans{(\delta,t)} s'$. It remains to prove that $\y'\lessdot_{Z\setminus U} \y$. Indeed,
    \begin{itemize}
        \item if $i \in U$ then moreover $i\in\reset_\delta$ and  $y_i=y'_i=0$;
        \item if $i \in Z\setminus U$ then either $y_i=y'_i=0$ or $y'_i=x'_i+t\leq x_i+t =y_i$, in both cases $y'_i\leq y_i$;
        \item if $i \not\in Z$ then $y'_i=x'_i+t= x_i+t =y_i$.
    \end{itemize}
   % \aldric{Or just say that time passing and reset are (component-wise)monotonic operations?\\}
    By definition, we conclude that $\y'\lessdot_{Z\setminus U} \y$. \qedhere
\end{itemize}
\end{proof}

\subsection{On speedy languages}
\lemspeedylight*
We will state and prove a more detailed version of this lemma, starting with the following:
\begin{con}\label{con:init}
In a speedy automaton $\D$, for every location $p$ let $\sigma_p$ be a path in $\D$ cycling from $p$ to $p$ and resetting all clocks in $\reset_\D$, and  $u_p$ the trace of $\sigma_p$ (timed word) of duration $0$. 
\end{con} 
%\aldric{Ce n'est pas une construction~?}

\begin{lemma}
\label{lem:fast:triv} The languages of $\D$ and its trivially timed form  $\widetilde{\D}$ are related as follows:
\begin{itemize}  
    \item for any states of the former automaton 
    $$
L_t^\D((p,\x),(q,\x'))\subseteq L_{t,\full}^{\widetilde{\D}};
$$
\item for every locations $p$ and $q$,  a clock valuation $\x\models S(p)$ and time $t$ satisfying $\x+t\models \guard_\D$, 
%there exists a  timed word $u_p$ of duration $0$ such that 
for $\y=(\x+t)[\reset_\D]$ 
it holds that
$$
 \overline{L_t^\D}((p,\x),(q,\y))\supset (u_p,0) \pql_{t}^{\widetilde{\D}}(u_q,t)
$$
with words $u_p$ and $u_q$ as in \cref{con:init}.
%\eugene{durée $=t$}
\end{itemize}
\end{lemma}

\begin{proof}
The former inclusion is evident.

For the latter, we first show that $$L_t^{\widetilde{\D}}((p,\x[\reset_\D]),(q,\y))=\overline{L_t^\D}((p,\x[\reset_\D]),(q,\y)).$$

First, observe that any intermediate state of any run $\rho$ of $\widetilde{\D}$ starting in $(p,S(p)[\reset_\D]$, ending in $q$ and of duration $t<1$ have clock vectors that satisfy $\guard_\D$. Indeed:\begin{itemize}
    \item For clocks in $\reset_\D$: since they all start with value $0$ and the run lasts $t$, their values stay in $[0,t]\subseteq [0,1]$. 
    \item For clocks not in $\reset_\D$: they may only grow from the initial state to the final state of $\rho$. But the clock vectors of both states, being resp. in $S(p)$ and $S(q)$, both satisfy $\guard_\D$, which is convex, in particular on coordinates outside $\reset_\D$.
\end{itemize}
Hence, all intermediate states of $\rho$ satisfy $\guard_\D$.

Then, we show that $\rho$ is actually a run of $\D$. We proceed by induction on prefixes of $\rho$: initially $q_0=(p,\x[\reset_\D])$ is a valid state of $\D$; assuming $\rho_n$, the $n$-th prefix of $\rho$, is a valid run of $\D$, then the next transition taken in $\widetilde\D$ is fired for a clock value satisfying $\guard_D$, obtained by elapsing time from $q_n$, the destination of the $\rho_n$, hence the corresponding transition in $\D$ could have been fired, leading to the same state $q_{n+1}$, which is thus valid state of $\D$, so $\rho_{n+1}$ is a valid run of $\D$.
So we got:
\begin{align*}
L_t^{\widetilde{\D}}((p,\x[\reset_\D]),(q,\y))&=\overline{L_t^\D}((p,\x[\reset_\D]),(q,\y));\\
(u_p,0)L_t^{\widetilde{\D}}((p,\x[\reset_\D]),(q,\y))(u_q,t)&=(u_p,0)\overline{L_t^\D}((p,\x[\reset_\D]),(q,\y))(u_q,t)\\&\qquad\qquad
\subseteq\overline{L_t^\D}((p,\x),(q,\y)).\qquad\qquad  \qed
\end{align*}
%$$L_t^{\widetilde{\D}}((p,\x[\reset_\D]),(q,\y))\subseteq\overline{L_t^\D}((p,\x),(q,\y)).$$
%
%Since the duration of $u_p$ and $u_q$ is $0$, and they are traces of $\sigma_p$ and $\sigma_q$ (with their known origin and destination), we have 
%$$
% (u_p,0) L_t^\D((p,\x[\reset_\D]),(q,\y))(u_q,t)\subseteq \overline{L_t^\D}((p,\x[\reset_\D]),(q,\y))(u_q,t)\subseteq
% \overline{L_t^\D}((p,\x),(q,\y)).   
%$$
%If we add that $\pql_t^{\widetilde{\D}}=L_t^\D((p,\x[\reset_\D]),(q,\y))$, we obtain the desired result.
\end{proof}

\subsection{Properties of abstraction}
\subsubsection{Time-divergence of $\widehat{\aut}$} follows from the statement below, corollary of \cref{lem:black:non0}. 
\begin{lemma}\label{lem:non:obese}
%$\widehat\aut$ is not obese, and
Every corner-point cycle in $\widehat\aut$ takes at least $1$ time unit.
\end{lemma}

\subsubsection{Quantitative properties.}
For proof needs, we augment the WTG from \cref{con:abstract} with transition labels (and thus obtain a weighted TA with the same reward-to-time ratios). Abusively, we use the same notation  $\widehat{\aut}$. 
\begin{con}
 The stratified automaton $\aut=(Q,X,\Pset{\Sigma},\Delta,S,I,F)$,  is transformed into the weighted TA
$\widehat{\aut}=(\widehat{Q},X,\widehat{\Sigma},\widehat{\Delta},\widehat{S},w)$ as in \cref{con:abstract} with the following differences:
\begin{itemize}
 
    \item $\widehat\Sigma=\Pset{\Sigma}\cup \{\pqd : \D\in \Dset \land p,q \in \D \}$, that is, we add to the alphabet an abstract letter for each spot and pair of its states.
  %  $\widehat(\Sigma)=\Sigma\cup\Dset$, that is we add to the alphabet an abstract letter for each spot.
    \item $\widehat\Delta$ is constructed as follows: 
       \begin{itemize}
           \item all the transitions in $\Delta$  except red ones within the same spot are preserved; 
           \item for a red transition $p\trans{a,\guard,\reset}q$ with $p\not\in \D$ and $q\in \D$  for some spot $\D$, as well as for  a black transition; $p\trans{a,\guard,\reset}q$ with  $q\in \D$, a copy  $p \trans{a,\guard,\reset}\check{q}$ is created; 
           %\item similarly for  each black transition $p\trans{a,\guard,\reset}q$ with  $q\in \D$   a copy  $p \trans{a,\guard,\reset}\check{q}$ is created;
           \item for each spot $\D$ and   locations $p,q\in \D$, a new abstract transition $\check{p} \trans{\pqd,\guard_\D,\reset_\D}q$ is introduced, with $\guard_\D$ and $\reset_\D$ as in \cref{def:speedy}.
       \end{itemize}
\end{itemize}
\end{con}
We will establish a simulation relation $\ll $(in some weak sense defined below) between $\hat{\aut}$
and $\aut$ as follows. Whenever $(p,\x)\lessdot (p,\x')$ in $\aut$, by  definition  $(p,\x)\ll (p,\x') $ and $(\check{p},\x)\ll (p,\x')$ (if $\check p$ exists).

\begin{lemma}\label{lem:sim:abs}
$\hat\aut$ simulates $\aut$ in the following sense.
\begin{itemize}
    \item Suppose $(r,\hat \x)\ll (r,\x)$. If $(r,\x)\Trans{u}(s,\y)$ in $\aut$ and the run does not involve any red transition within any spot, then there exists $ (s,\hat \y) \ll (s,\y)$ and a run  $(r,\hat \x)\Trans{u}(s,\hat \y)$ in $\hat{\aut}$. Moreover, if there exists a location $\check s$, then there exists another run $(r,\hat \x)\Trans{u}(\check s,\hat \y)$
    \item Suppose $(\check s,\hat \x)\ll (s,\x)$. If $(s,\x)\Trans{v}(r,\y)$ in $\aut$  and the run only involves red transitions within a spot, then there exists $(r,\hat \y) \ll (r,\y)$ and a transition of the form $(\check s,\hat \x)\trans{t,\pqd}( r,\hat \y)$ in $\hat{\aut}$, where $t=\length(v)$.
\end{itemize}
\end{lemma}
\begin{proof}
For the former case, we first build a run in $\aut$  (which must exist since $\lessdot$ is a simulation) and next reproduce it in $\hat\aut$ (which is possible since it does not involve any red transition within a spot. Redirecting the last transition towards $\check Q$ we obtain another run to $(\check s,\hat \y)$.

For the latter case, since  $(s,\x)\Trans{v}(r,\y)$  takes place in some spot, it necessarily goes from an avatar $s=(p,Z)$ to an avatar $r=(q,Z)$ (with the same $Z$ since it is the same spot). The set $\reset$ of clocks reset in $\D$ is a subset of $Z$. Thus  $y_i=x_i+t$ for all $i\not\in Z$.
On the other hand, we have $\hat{\x} \lessdot_Z \x$, and 
in $\hat{\aut}$, by construction,  there is a transition $(\check{s},\hat{\x}) \trans{t,\pqd}(r,\hat{\y})$ with $\hat{\y}=(\hat {\x}+t)[\reset_\D]$. Let us compare $\hat{\y}$ and $\y$: 
\begin{itemize}
    \item for all $i\in \reset_\D$, it holds that $\hat{y}_i=0\leq y_i$;
    \item for all $i\in Z\setminus \reset_\D$, it holds that $\hat{y}_i=\hat{x}_i+t\lessdot x_i+t=y_i$;
    \item for all $i \not\in Z$, it holds that $\hat{y}_i=\hat{x}_i+t= x_i+t=y_i$.
\end{itemize}
We conclude that $\hat{\y}\lessdot_Z \y$ and thus $(r,\hat \y) \ll (r,\y)$. \qed
\end{proof}

\begin{definition}[approximating functions]
   Let $w$ be a timed word in $\widehat\Sigma$. We define $\overline\gamma(w)$ as the language obtained by replacing each abstract letter $(\pqd,t)$ by the trivially timed language $L_{\full,t}^{\aut{\D}}$. Similarly $\underline\gamma(w)$ is the language obtained by replacing each abstract letter $(\pqd,t)$ by the  timed language $(u_p,0)\pql_t(\aut{\D})(u_q,t)$, as in \cref{lem:fast:triv}.
\end{definition}

\begin{lemma}\label{lem:approx:hat:over}
Overapproximation property of $\widehat{\aut}$ and $\overline{\gamma}$:
$$
 L_T(\aut) \subseteq U_1\cdot \overline\gamma(L_{\full,T}(\widehat\aut))\cdot  U_1,  
$$
where $U_1$ is the universal language of all timed words on $\Sigma$ of duration $\leq 1$. 
\end{lemma}
\begin{proof}
For the first inclusion, consider a run  $\rho$ of $\aut'$ on some timed word $w$. Let $\rho_1$ be the maximal red prefix, $\rho_3$ the maximal red suffix, and $\rho=\rho_1\rho_2\rho_3$, let $w=w_1w_2w_3$ be the corresponding factorization of $w$. Due to \cref{lem:heart}, $\length(w_1),\length(w_3)<1$, and these two words necessarily belong to $U_1$. It remains to prove that $w_2$, trace of the run $\rho_2$ starting and ending by black transitions, belongs to $\overline\gamma(L_t(\widehat\aut))$.

To that aim, we partition $\rho_2$ into fragments:
$$\rho_2 = r_0 \Trans{u_0} s_1 \Trans{v_1}r_1\Trans{u_1} s_2 \Trans{v_2} r_2\cdots \Trans{u_n}f,
$$
where the fragment from $s_i$ to $r_i$ happens within a spot $\D_i$, while that from $r_i$ to $s_{i+1}$ outside of spots.
We remark that $w=u_0v_0u_1v_1\dots u_n$.

We will build a run of $\hat\aut$ of the form
$$\hat\rho_2 = \hat{r}_0 \Trans{u_0} \hat{s}_1 \trans{ t_1,\pqd_1}\hat{r}_1\Trans{u_1} \hat{s}_2 \trans{t_2,\pqd_2}\hat{r}_2\cdots \Trans{u_n}\hat f,
$$
where all $s_i$ are abstract state
$$
f, \hat{r}_i\in Q; \quad \hat{r}_i\ll r_i;\quad \hat{s}_i\ll s_i;\quad t_i=\length(v_i).
$$
To that aim, we start in $\hat{r}_0=r_0$ and apply \cref{lem:sim:abs} inductively to every fragment $u_i$ and $v_i$.

We take the run $\hat\rho_2$ and apply to its trace $\hat{w}\in L_{\full, T}(\hat A)$ the operator $\overline{\gamma}$, which gives 
$$
\overline{\gamma}(\hat w)= u_0 L_{\full,t_1}^{\D_1} u_1 L_{\full,t_2}^{\D_2}\cdots u_n \ni u_0 v_1 u_1 v_2 \cdots u_n=w_2,
$$
which concludes the proof of the first inclusion. \qed
\end{proof}

\begin{lemma}\label{lem:approx:hat:under}
Underapproximation property of $\widehat{\aut}$ and $\underline{\gamma}$ for black states $r,f$:
$$
\overline{\rfl_T(\aut')}\supset \overline{\underline\gamma( \rfl_T(\widehat{\aut})}). 
$$
\end{lemma}
\begin{proof}
Given a run of $\hat\aut$ on some word $w$ of the form
$$\hat\rho_2 = r_0 \Trans{u_0} {s}_1 \trans{ t_1,\pqd_1}{r}_1\Trans{u_1} {s}_2 \trans{t_2,\pqd_2}{r}_2\cdots \Trans{u_n}f,
$$
each $r_i \Trans{u_i} {s}_{i+1}$ can be reproduced in $\aut'$, and for each transition ${s}_1 \trans{ t_1,\pqd_1}{r}_1$, the whole language $ \pql_t^{\D}$ has runs from $s_i$ to $r_i$ in $\overline\aut'$. Consequently $\underline{\gamma}(w)\subseteq \overline{\rfl_T(\aut')}$ as required.  \qed
\end{proof}

\section{Proof of Main result }
We split the proof into that of the lower and the upper bound.

\subsection{Lower bound}

\begin{lemma} \label{lem:lower}
$\alpha/\varepsilon \lessapprox \band\capa_\varepsilon(\aut).$
\end{lemma} 
\begin{proof}
If $\alpha=0$, there is nothing to prove. Hereinafter $\alpha>0$. 
By \cite{alive}, there exists a corner-point cycle $\sigma_0$ in $\widehat{\aut}$  on some timed word $v$ with weight-to-time ratio $\alpha$, let its duration be $t_0$, the number of abstract letters $n_0$ and weight $W(v)=\alpha t_0$. We remark that $\sigma_0$ spends in each state either $0$ or $1$  time unit. Let $\varepsilon <1/4$. By construction of the weighted TA $\widehat\aut$,
$$
\alpha t_0=W(v)=\sum_{(\pqd,1) \text { in }v}\alpha_D.
$$

For any duration $T$, let $K=\lfloor T/t_0\rfloor$, and consider the timed word $v^K$, its duration is in $(T-t_0,T]$.

We build an $\large\varepsilon$-separated set $\Sep$ in $\underline\gamma (v^K)$ by replacing each factor $ \pql_1(\widetilde{\D})$ (corresponding to an abstract letter $(\pqd,t)$ with $t=1$ )  by  $\Sep_{1,\varepsilon}^{pq}$ from \cref{prop:trivial:bis}. As for letters  $(\pqd,0)$, we replace them by any timed word in $\pql_0^{\widetilde{\\D}}$. 

By construction $\Sep \subseteq {\underline\gamma(v^K)}$ and thus, by \cref{lem:approx:hat:under}, $\Sep\subseteq \overline \rrl_T(\aut)$ for some corner-point $r$, and it is easy to see that $\Sep$ is indeed $\varepsilon$-separated. Taking timed words $u$ leading from $I$ to $r$ and $v$ from $r$ to $F$ we get an $\varepsilon$-separated $u\cdot\Sep\cdot v\subseteq \overline{L_{T+\length(uv)}(\aut)}$ of the same cardinality.

It remains to estimate its cardinality using the lower bound from \cref{prop:trivial:bis}
% \begin{multline*}
%     \log\# \Sep\geq K\sum_{(\pqd,t) \text { in }v}\left(\log c +(1-\varkappa)\alpha_D t/\varepsilon\right) \geq\\ \frac{T-t_0}{t_0\varepsilon}\cdot \left(n_0\log c +(1-\varkappa)\sum_{(\pqd,t) \text { in }v}\alpha_D t\right)\geq T(1-2\varkappa) \alpha/\varepsilon,    
% \end{multline*}
%
$$
    \log\# \Sep= K\sum_{(\pqd,1) \text { in }v} \#\Sep_{1,\varepsilon}^{pq}\gtrapprox K\sum_{(\pqd,1) \text { in }v} \frac{\alpha_D}{\varepsilon} -KC=\frac{Kt_0\alpha}{\varepsilon}-KC,
$$
and thus
$$
\lim_{T\to\infty} \frac{\log\# \Sep}{T}\gtrapprox\frac{\alpha}{\varepsilon} -\frac{C}{t_0}\eqapprox {\alpha}/{\varepsilon},
$$
as required. 

We conclude that the entropic bandwidth of $\overline{L(\aut)}$ is at least $\alpha/\varepsilon$ and use \cref{cor:closed} to prove the same for $L(\aut)$ and $\varepsilon'=\varepsilon/(1+\varkappa)$, which implies the asymptotic inequality to prove. \qed
\end{proof}

\subsection{Upper bound}
\begin{lemma}
If $\aut$ is  obese then $\alpha>0$.
\end{lemma}
\begin{proof}
Suppose an \CTA\ $\aut$ is obese. Its stratified version $\aut'$ is also obese since it accepts the same timed language.  
By \cite[Thm. 4]{3classes}, it contains an obesity pattern. From this pattern, we will be able to extract a path of non-zero weight in the weighted timed-graph, and thus deduce that $\alpha>0$. We proceed as follows:
 
 The obesity pattern states that an obese \RTA\ has a location $q$ with two cycles $\pi_f$ and $\pi_r$ from $q$ such that in the CPA there exists, for some $u$ and $v$, corner-points of location $q$:
 \begin{itemize}
     \item a cycle along $\pi_f$, from $u$ to $u$, of total duration $0$;
     \item a cycle along $\pi_f$, from $v$ to $v$, of total duration $0$;
     \item a path along $\pi_f$, from $u$ to $v$, of total duration $\geq 1$;
     \item a path $\sigma_r$ along $\pi_r$, from $v$ to $u$.
 \end{itemize}
 
Observe that due to \cref{lem:black:non0}, $\pi_f$ is fast, hence all edges in $\pi_f$ are red, and by \cref{lem:heart} applied to the stratified automaton,  all edges of $\pi_f$ are contained in the same spot $\D$.

Consider now $\sigma_r$.
%, the path along $\pi_r$, going from $v$ to $u$. 
This path has to leave the spot $\D$ at some point (otherwise, by going from $u$ to $v$ in $\geq 1$ time units, then back to $u$ and back to $v$ in $\geq 1$ time units, more than one time unit would be spent on the same spot). Let $v'$ be the last corner-point in $\sigma_r$ before leaving $\D$ for the first time, and let $u'$ be the first corner-point after re-entering $\D$ for the last time. Let $p$ and $q$ be the locations corresponding to $u'$ and $v'$, respectively.

Note that there are paths from $u'$ to $u$ and from $v$ to $v'$ (they have to be of duration $0$ to respect the limit of one time unit).

By construction, in the trim abstracted automaton $\hat{\aut}$, there is an edge from $\check{p}$ to $q$ labeled by $\pqd$. In the corresponding weighted corner-point graph, there is an edge from $u'$ to $v'$ with reward $\alpha_\D$ and timing $1$, which is moreover part of a cycle. It suffices now to show that $\alpha_\D>0$, which follows from \cref{prop:SCC:squeez}, item 1. \qed
\end{proof}

\begin{lemma}
$\band\ent_{\varepsilon/2}(\aut)\lessapprox \alpha/\varepsilon$ if $\alpha >0$.
\end{lemma}
\begin{proof}
We build an $\varepsilon/2$-net for $\overline\gamma(L_T(\widehat\aut))$ as follows. 

By \cref{lem:non:obese} $\widehat\aut$ is not obese, and thus $L_T(\widehat\aut)$  has bandwidth $O(\log 1/\varepsilon)$ and thus admits an $\varepsilon/2$-net $\widehat{\Net}_{T,\abs}$ of size $\leq\varepsilon^{-CT}$ (for $\varepsilon$ small enough). Also, the number of letters in a word of duration $T$ is at most $CT+C$.
Due to \cref{lem:alive}, the weight of each timed word $v\in\widehat{\Net}_{\abs}$  can be bounded as follows:  
$
W(v)\lessapprox\alpha T. %+C;
$

We build a timed language $\Net(v)$ replacing every abstract letter $(\pqd,t)$ by $\Net_{t,\varepsilon}$ as defined in \cref{prop:trivial:bis}. Then we make a union $\Net_T=\bigcup_{v\in\widehat{\Net}_\abs}  \Net(v)$. It is easy to see that $\Net_T$ is an  $\varepsilon/2$-net for $\overline\gamma(L_T(\widehat\aut))$.%\eugene{Bernardo, est-ce vra? doit-on changer légèrement $\varepsilon$, faut-il ajouter un lemme sur concaténation de langages au début?}

To estimate its cardinality we first estimate $\#\Net(v)$ using \cref{prop:trivial:bis}: 
$$
    \log\#\Net(v)\lessapprox\sum_{(\pqd,t) \text { in } v} \frac{\alpha_D t}{\varepsilon}
    \lessapprox(CT+C)C +  \frac{W(v)}{\varepsilon}  \lessapprox\frac{\alpha T+C'}{\varepsilon} +C'T, 
$$
with some constant $C'$. Knowing that $$\Net=\bigcup_{v\in \widehat{\Net}_{\abs} }\Net(v),$$
we obtain:
$$
\log \#\Net\lessapprox \log\#\widehat{\Net}_{\abs}+\frac{\alpha T+C'}{\varepsilon} +C'T\leq -C'T\log \varepsilon +\frac{C'}{\varepsilon}  + \frac{\alpha T}{\varepsilon },
$$
and thus
$$
\bandh_{\varepsilon/2}(\aut)= \limsup_{T\to\infty} \log \frac{\#\Net}T\lessapprox \frac{\alpha}{\varepsilon}-C'\log\varepsilon= \frac{\alpha(1-(C'/\alpha)\varepsilon\log\varepsilon)}{\varepsilon}\lessapprox \frac\alpha\varepsilon. \qedhere
$$
\end{proof}
\begin{figure}[H]
\begin{minipage}{0.5\textwidth}
\begin{lstlisting}[title={Bandwidth of obese timed  automata},label=list:band,mathescape=true]
real bandwidth(A: TimedAut)
    AH=addHeartAndUrgency(A)
    AR=regionSplit(AH)
    AZ=eliminate0(AR)
    red=detectRed(AZ)
    AS=stratify(AZ,red)
    $\Dset$=detectSpots(AS,red)
    for each ($\D$ in $\Dset$)
        alpha[$\D$]=growth($\D$)
    AA=abstract(AS,$\Dset$,alpha[])
    band=rewardPerTime(AA)
    return band
\end{lstlisting}
\end{minipage}
\hfill
\begin{minipage}{0.4\textwidth}
\begin{lstlisting}[title={Growth of speedy timed automata},label=list:growth,mathescape=true]
real growth(A: TimedAut)
    AU=support(A)
    ASq=squeeze(AU)
    AD=determinize(ASq)
    M=adjacencyMatrix(AD)
    rho=spectralRadius(M)
    return log(rho)
    
    
    
    
    $\phantom{a}$
\end{lstlisting}
\end{minipage}

\caption{Algorithm for computing the bandwidth of obese timed automata and its subroutine}\label{fig:algos}
\end{figure}

\begin{figure}[H]\small
\begin{center}
\begin{tikzpicture}[every node/.style={text=black}]
\node(prepxin2)  at (0,4)  {6};
\node(prepxin3)  at (1,4)  {7};
\draw (prepxin2) edge[thick,violet] (prepxin3);
\node[fit=(prepxin2)(prepxin3),rounded corners,thin,dashed,draw,fill opacity=0.2, fill=gray]{};

\node(pxin)  at (3,4)  {$\check p$};
\node[fit=(pxin),rounded corners,thin,dashed,draw]{};

\node(prepx0)  at (5,3.5)  {$\substack{0\\0}$};
\node(prepx1)  at (6,4.5)  {$\substack{1\\1}$};
\draw (prepx0) edge[thick,violet] (prepx1);
\node[rotate fit=45,fit=(prepx0)(prepx1),rounded corners,thin,dashed,draw,fill opacity=0.2, fill=gray]{};

\node(px0)  at (8,3.5)  {$0$};
\node(px1)  at (8,4.5)  {$1$};
\draw (px0) edge[thick,violet,right]  node {$p$}(px1);
\node[fit=(px0)(px1),rounded corners,thin,dashed,draw]{};

\node(preqin00) at (4,1.5) {$\substack{0\\0}$};
\node(preqin01) at (4,2.5) {$\substack{0\\1}$};
\node(preqin11) at (5,2.5) {$\substack{1\\1}$};
\draw[thick,violet] (preqin00)--(preqin01)--(preqin11) -- (preqin00);
\node[fit=(preqin00)(preqin01)(preqin11),rounded corners,thin,dashed,draw,fill opacity=0.2, fill=gray]{};

\node(qin0) at (8,1.5) {0};
\node(qin1) at (8,2.5) {1};
\draw (qin0) edge[thick,violet,right]  node {$\check q$}(qin1);
\node[fit=(qin0)(qin1),rounded corners,thin,dashed,draw]{};

\node(q00) at (11,1.5) {$\substack{0\\0}$};
\node(q01) at (11,2.5) {$\substack{0\\1}$};
\node(q11) at (12,2.5) {$\substack{1\\1}$};
\draw[thick,violet] (q00) -- (q01) -- (q11) -- (q00);
%\draw (q0) edge[thick,violet,above,near start] node {$q$} (q1);
\node[fit=(q00)(q01)(q11),rounded corners,thin,dashed,draw]{$q$};

\node(rin1) at (9,0) {1};
\node(rin2) at (10,0) {2};
\draw (rin1) edge[thick,violet,above,near end] node {$\check r$} (rin2);
\node[fit=(rin1)(rin2),rounded corners,thin,dashed,draw]{};

\node(r1) at (3,0) {1};
\node(r2) at (4,0) {2};
\draw (r1) edge[thick,violet,above,near end] node {$r$} (r2);
\node[fit=(r1)(r2),rounded corners,thin,dashed,draw]{};

\graph[edge quotes={fill=white,inner sep=1pt}]{
(prepxin2)->[bend left,thick,olive] (pxin)->["2.863/1",thick,teal] (prepx1) ->[thick,teal] (px1);
(prepxin3)-> (pxin) ->[thick,orange] (prepx0) ->[thick,orange] (px0) ;

(px0)->[pos=0.6,thick,orange] (preqin00)->[thick,orange] (qin0)-> (q00) ;
(qin0)->["2/1",thick,bend left=1.5cm,orange] (q11);
(px0)->["1"] (preqin11)-> (qin1);
(px1)->[thick,teal] (preqin01)->[bend right,thick,teal] (qin1)->[thick,teal] (q01) ;
(q00)->["1",bend right=6mm](rin1)->["3/1",thick,olive](r2);
(q01)->["1",thick,teal,bend right=10mm,near end](rin1);
(q00)->["2"](rin2);
(q01)->["2",near end,bend right=2mm](rin2);
(q11)->[near end,thick,orange,bend left=20mm](rin1)->[bend right](r1)->["5",near end] (prepxin2);
(r1)->["6"] (prepxin3);
(q11)->["1",near end,bend left =5mm](rin2)->[bend left=5mm](r2)->["4",bend left=2.5cm,thick,olive](prepxin2);
(r2)->["5",near start](prepxin3);
};
\end{tikzpicture}
\end{center}
\vspace*{-10mm}
\caption{Corner-point graph for the running example, see \cref{fig:avatars}, right (0s are omitted, one number on an arrow denotes a time, two numbers --- a reward and a time, gray rectangles correspond to regions after time elapse but before the clock reset); its optimal olive and teal cycle with ratio $\alpha\approx 0.838$; its suboptimal olive and orange cycle with ratio $\approx 0.833$.\label{fig:corners}
}\end{figure}
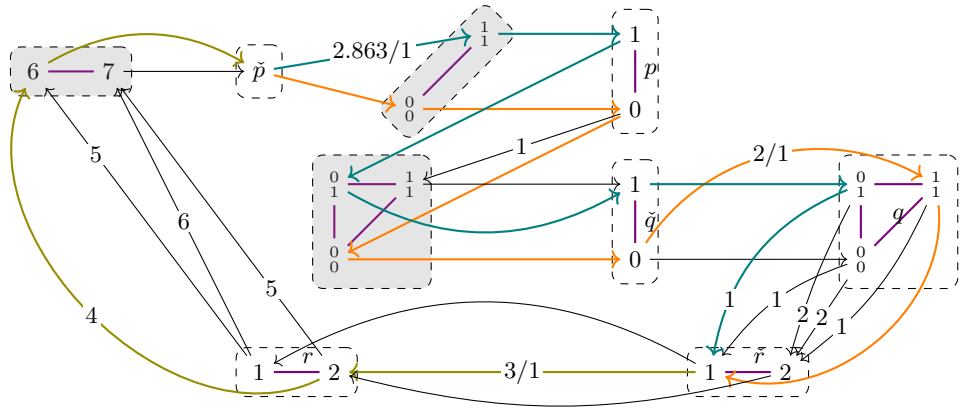
\newpage

\section{Summarizing the algorithm and finalizing the running example}
We present the overall algorithm for computing the bandwidth of obese timed automata on \cref{fig:algos}.

Let us return to the example from \cref{sec:example,fig:avatars}, for which  we have reduced the bandwidth computation to the optimal reward-per-time ratio for the WTG on the right of that figure. \cref{fig:corners} illustrates the computation of this ratio, using the method from \cite{alive}.  In the corner-point graph, the violet edges correspond to possible values of the clock $x$. Every graph edge is labeled by its reward and time. The optimal cycle of duration $7$ is highlighted in olive and teal, its ratio, and the automaton bandwidth coefficient is 
$\left(3+\log((7 + \sqrt{57})/2)\right)/{7}\approx 0.838$. Another interesting cycle, olive and orange, is not as good; its ratio is only $5/6\approx 0.833$.
\end{document}